\documentclass[12pt,a4paper]{article}
\usepackage[pdftex]{graphicx}
\pdfoutput=1 % needed for arXiV
\graphicspath{{./img/}}%Path relative to the main .tex file
 
\usepackage{jheppub}

\usepackage{subcaption}
\usepackage{xspace}
\usepackage{todonotes}
\usepackage{epigraph}
\usepackage{subcaption}
\usepackage{mwe}

\usepackage[utf8]{inputenc}

\usepackage[style=numeric-comp,sorting=none,backend=biber]{biblatex} 

\abstract{In modern High Energy Physics (HEP) experiments visualization of experimental data has a key role in many activities
and tasks across the whole data chain: from detector development to monitoring, from event generation to reconstruction of physics objects,
from detector simulation to data analysis, and all the way to outreach and education.
In this paper the definition, status, and evolution of data visualization for HEP experiments will be presented.
Suggestions for the upgrade of data visualization tools and techniques in current experiments will be outlined, along with
guidelines for future experiments. This paper expands on the summary content published in the
HSF \emph{Roadmap} Community White Paper~\cite{HSF-CWP-2017-01}.}

\begin{document}

\noindent
\begin{tabular*}{\linewidth}{lc@{\extracolsep{\fill}}r@{\extracolsep{0pt}}}
& & HSF-CWP-2017-15 \\
& & October 26, 2018 \\ % use \date or hardwire e.g. December 15, 2017
& & \\
\end{tabular*}
\vspace{2.0cm}

\title{HEP Software Foundation Community White Paper Working Group -- Visualization}

\author{HEP Software Foundation:}

\author[a,b]{Matthew Bellis}
\author[c,1]{Riccardo Maria Bianchi\note{Paper Editors}}%this same note will be shared with all the other authors with the [1] (i.e. Thomas McCauley)
\author[d]{Sebastien Binet}
\author[e]{Ciril Bohak}
\author[f]{Benjamin Couturier}
\author[g]{Hadrien Grasland}
\author[h]{Oliver Gutsche}
\author[i]{Sergey Linev}
\author[j]{Alex Martyniuk}
\author[k,1]{Thomas McCauley}
\author[l]{Edward Moyse}
\author[m]{Alja Mrak Tadel}
\author[n]{Mark Neubauer}
\author[f]{Jeremi Niedziela}
\author[p]{Leo Piilonen}
\author[q]{Jim Pivarski}
\author[r]{Martin Ritter}
\author[s]{Tai Sakuma}
\author[m]{Matevz Tadel}
\author[f]{Barthélémy von Haller}
\author[t]{Ilija Vukotic}
\author[j]{Ben Waugh}

\affiliation[a]{Siena College, Loudonville NY, USA}
\affiliation[b]{Cornell University, Ithaca NY, USA}
\affiliation[c]{University of Pittsburgh, Pittsburgh PA, USA}
\affiliation[d]{CNRS/IN2P3, Clermont-Ferrand, France}%{LSST, Alice, LPC}
\affiliation[e]{University of Ljubljana, Ljubljana, Slovenia}
\affiliation[f]{CERN, Geneva, Switzerland}
\affiliation[g]{LAL, Université Paris-Sud and CNRS/IN2P3, Orsay, France}
\affiliation[h]{FNAL, Batavia IL, USA}
\affiliation[i]{GSI Darmstadt, Germany}
\affiliation[j]{University College London, London, UK}
\affiliation[k]{University of Notre Dame, Notre Dame IN, USA}
\affiliation[l]{University of Massachusetts, Amherst MA, USA}
\affiliation[m]{University of California at San Diego, San Diego CA, USA}
\affiliation[n]{University of Illinois, IL, USA}
\affiliation[p]{Virginia Tech, VA, USA}
\affiliation[q]{Princeton University, Princeton PA, USA}
\affiliation[r]{LMU Munich, Munich, Germany}
\affiliation[s]{University of Bristol, Bristol, UK}
\affiliation[t]{University of Chicago, Chicago IL, USA}

\maketitle

\newpage

\hypertarget{scope}{%
\section{Scope}\label{scope}}

\epigraph{Visualization: Turning numbers into pixels}{\textit{Hadrien Grasland \\ HSF Workshop 2017, Annecy}}

This paper will describe three kinds of data visualization used in High-Energy Physics (HEP): interactive visualization of event data in
applications known commonly as event displays, statistical data visualization such as histograms,
and non-spatial data visualization such as networks and graphs.

Event displays are the main tool used to explore experimental data at the event level.
There are two main types of displays. The first type are those that are integrated into an experiment's software frameworks,
which are usually able to access and visualize all experimental data at the cost of greater application complexity and lesser portability.
The second type of displays are those designed as cross-platform applications, lightweight and fast, often
delivering a simplified version or a subset of the event data. All event displays show the detector geometry; the level of detail
displayed depends on the application's use-case and targeted audience as well as on the application's capability to
render geometries responsively.

Beyond event displays, HEP also uses statistical data visualizations such as histograms, which display the distributions of data variables
in aggregate over multiple events. These visualizations are not strongly linked to a detector geometry. Data
analysis tools and techniques used in HEP are
described in the HSF \textit{Data Analysis and Interpretation} Community White Paper~\cite{HSF-CWP-2017-05}.

The final types of visualization considered in this paper are those that visualize non-spatial data, such as the graphs used to
visually describe the structure of the detector description, that is
the representation of all geometrical volumes that compose the sub-detectors and the infrastructure of a HEP experiment.
More details about the detector geometry can be found in the HSF \textit{Detector Simulation} Community White Paper~\cite{HSF-CWP-2017-07}.

Other types of data visualization used in HEP experiments, such as visualization for slow control
or dashboards for data analytics, are considered out of scope and are not discussed.

The content of this paper is the summary of the direct experience of the authors in designing, building, and deploying
interactive data visualization applications for the experiments ALICE, ATLAS, Belle II, CMS, LHCb, and LSST, for scientific
software frameworks such as ROOT, and in other cross-experiment projects. It is the outcome of a number of discussions and workshops organized
under the auspices of the HEP Software Foundation,
in which the authors worked to build a common, shared view of the field, its current issues, and the potential ways it could and should evolve
in the context of HEP.

\hypertarget{current-landscape}{%
\section{Current landscape}\label{current-landscape}}

\hypertarget{event-displays}{%
\subsection{Event displays}\label{event-displays}}

Three key features characterize HEP event displays. The first is an \emph{event-based workflow}. Applications access
experimental data on an event-by-event basis,
visualizing the data collections belonging to a particular event. Data can be related to the
actual physics events ({\it e.g.}\ a collection of reconstructed physics objects, like jets and tracks) or to the experimental
conditions ({\it e.g.}\ different versions of the detector description and calibration data).

The second key feature is \emph{geometry visualization}.
The level of geometric detail displayed depends on the specific use-case, on the way the geometry information is stored and
fetched ({\it e.g.}\ from a database as part of a software framework or from an external file), and on
limitations of the application itself along with
considerations about speed, efficiency, and portability.

The third key feature is \emph{interactivity}. Applications offer different interfaces and tools for users to interact
with the visualization, select event data, and to set cuts on objects' properties.
In addition to the interactive usage, applications often store different settings to automate user actions.

\begin{figure*}
	\centering
	\begin{subfigure}[b]{0.475\textwidth}
		\centering
		\includegraphics[width=\textwidth]{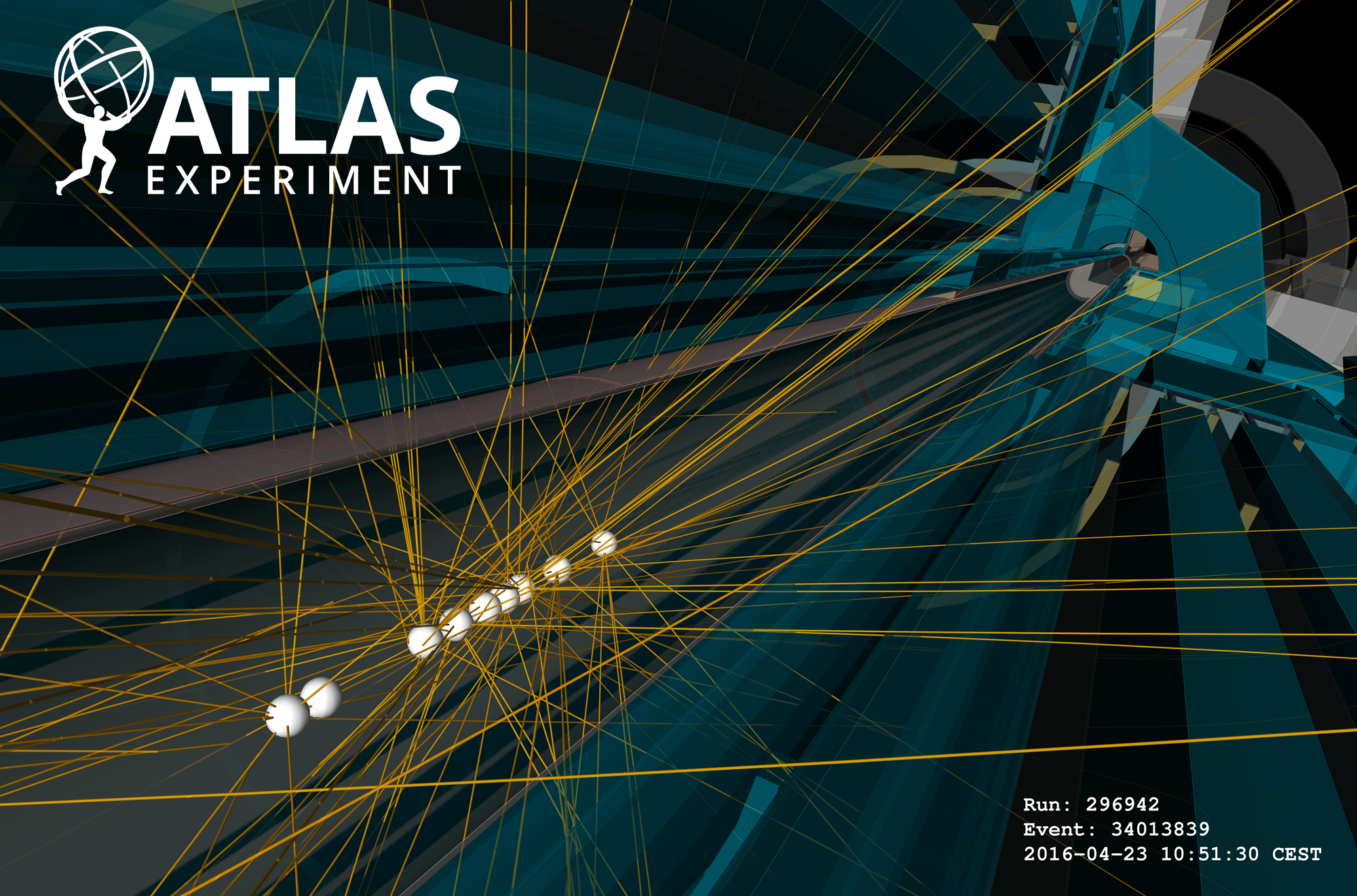}
		\caption[ATLAS 3D event display]%
		{{\small ATLAS}}
		\label{fig:ed-atlas}
	\end{subfigure}
	\quad
	\begin{subfigure}[b]{0.475\textwidth}
		\centering
		\includegraphics[width=\textwidth]{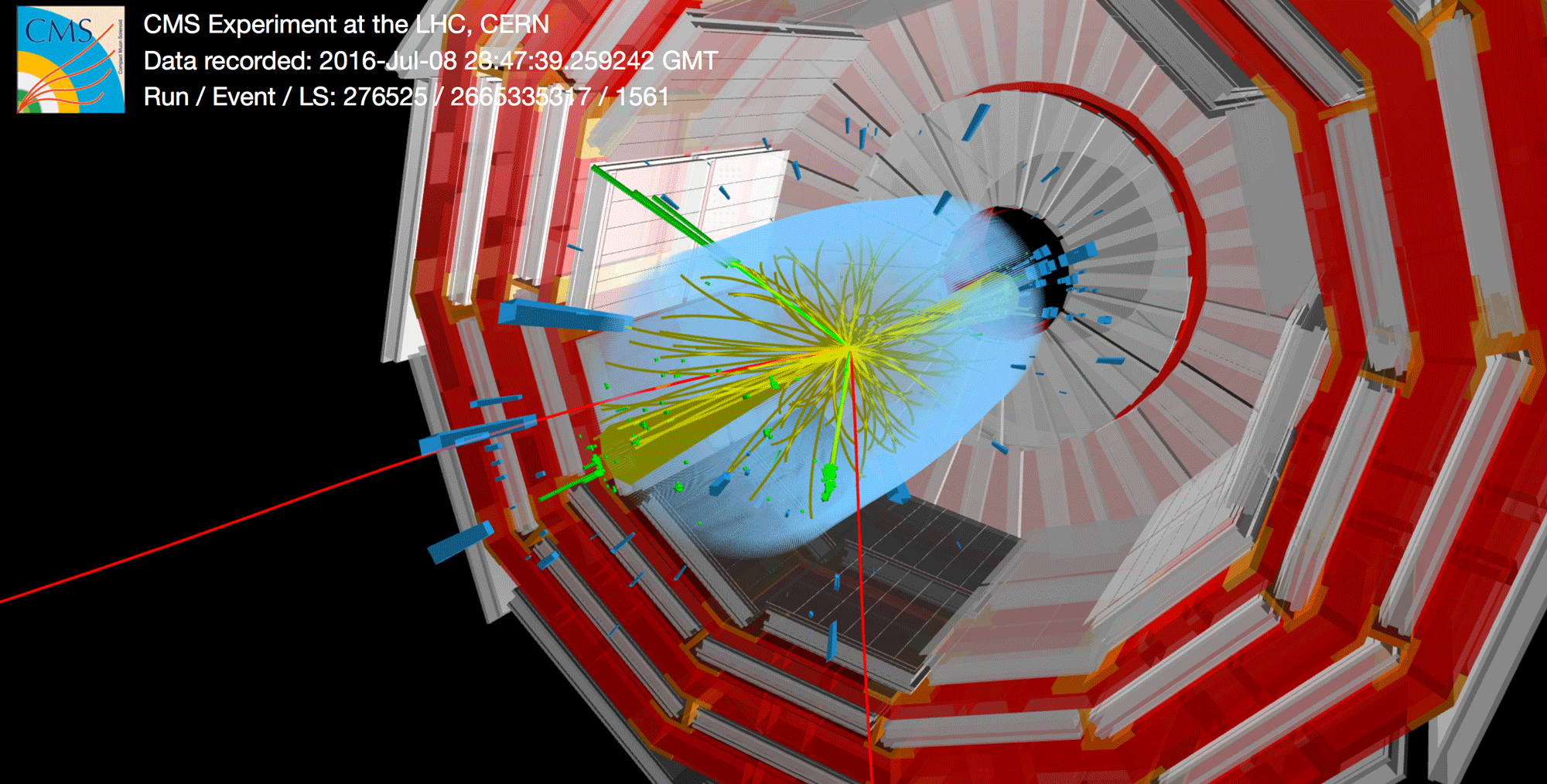}
		\caption[CMS event display]%
		{{\small CMS}}
		\label{fig:ed-cms}
	\end{subfigure}
	\vskip\baselineskip
	\begin{subfigure}[b]{0.475\textwidth}
		\centering
		\includegraphics[width=\textwidth]{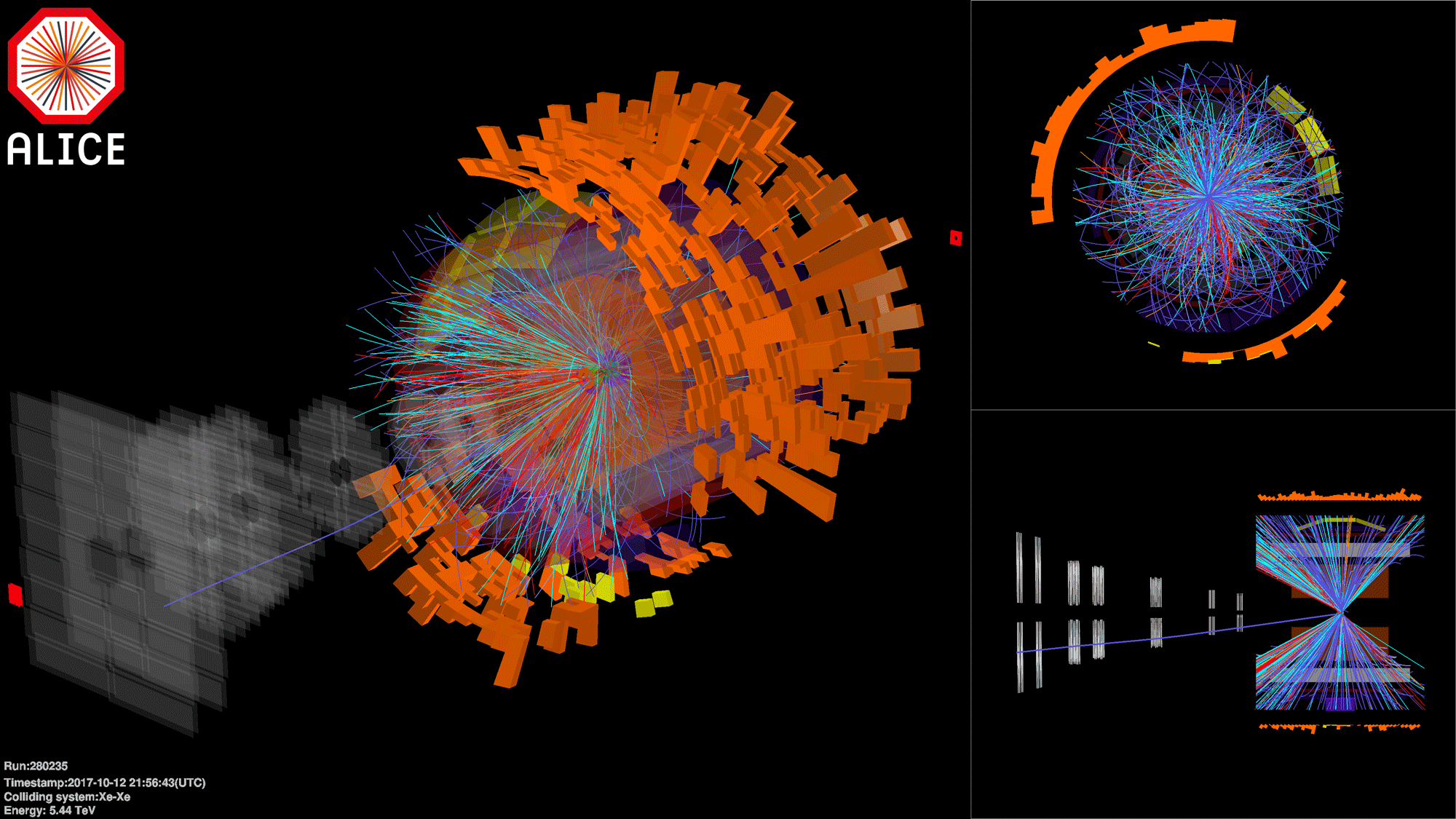}
		\caption[ALICE event display]%
		{{\small ALICE}}
		\label{fig:ed-alice}
	\end{subfigure}
	\quad
	\begin{subfigure}[b]{0.475\textwidth}
		\centering
		\includegraphics[width=\textwidth]{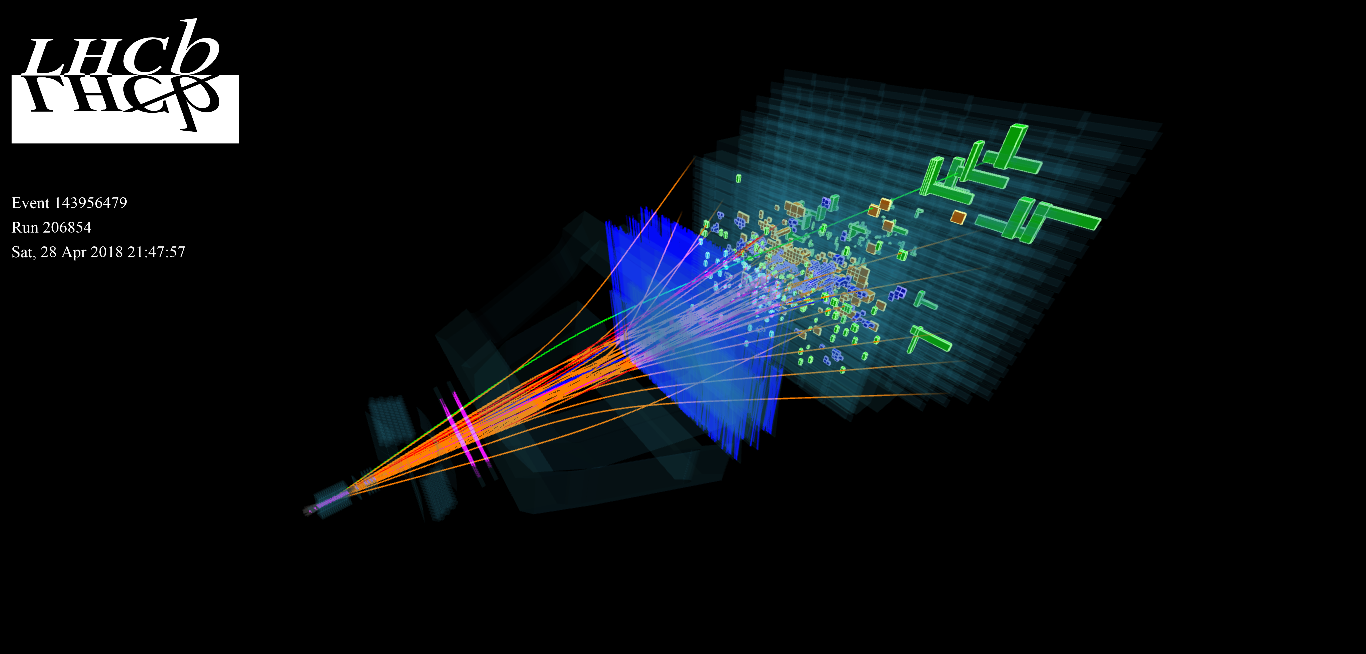}
		\caption[LHCb event display]%
		{{\small LHCb}}
		\label{fig:ed-lhcb}
	\end{subfigure}
	\vskip\baselineskip
	\begin{subfigure}[b]{0.475\textwidth}
		\centering
		\includegraphics[width=\textwidth]{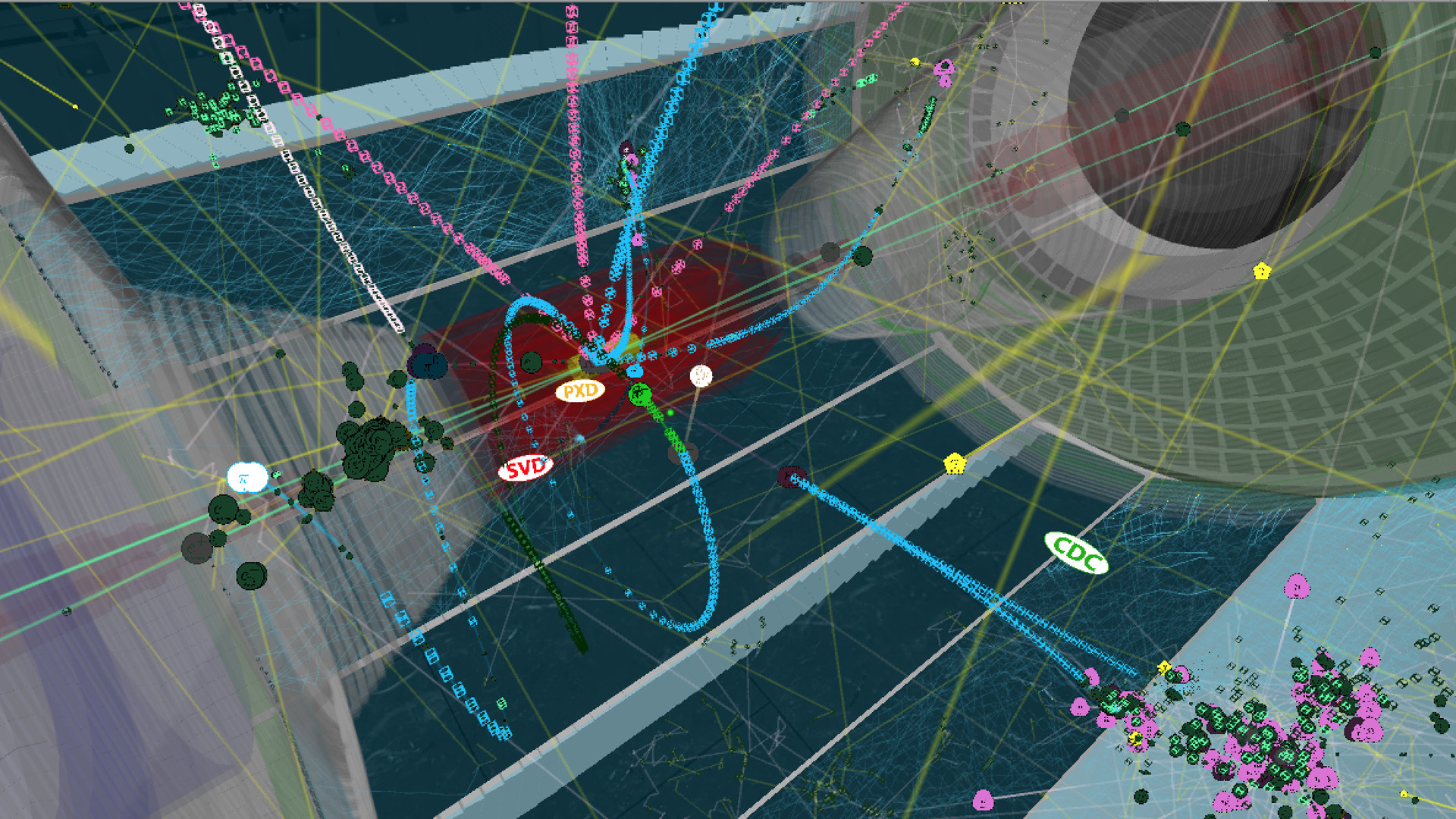}
		\caption[Belle II VR event display]%
		{{\small Belle II VR}}
		\label{fig:ed-belleii}
	\end{subfigure}
	\quad
	\begin{subfigure}[b]{0.475\textwidth}
		\centering
		\includegraphics[width=\textwidth]{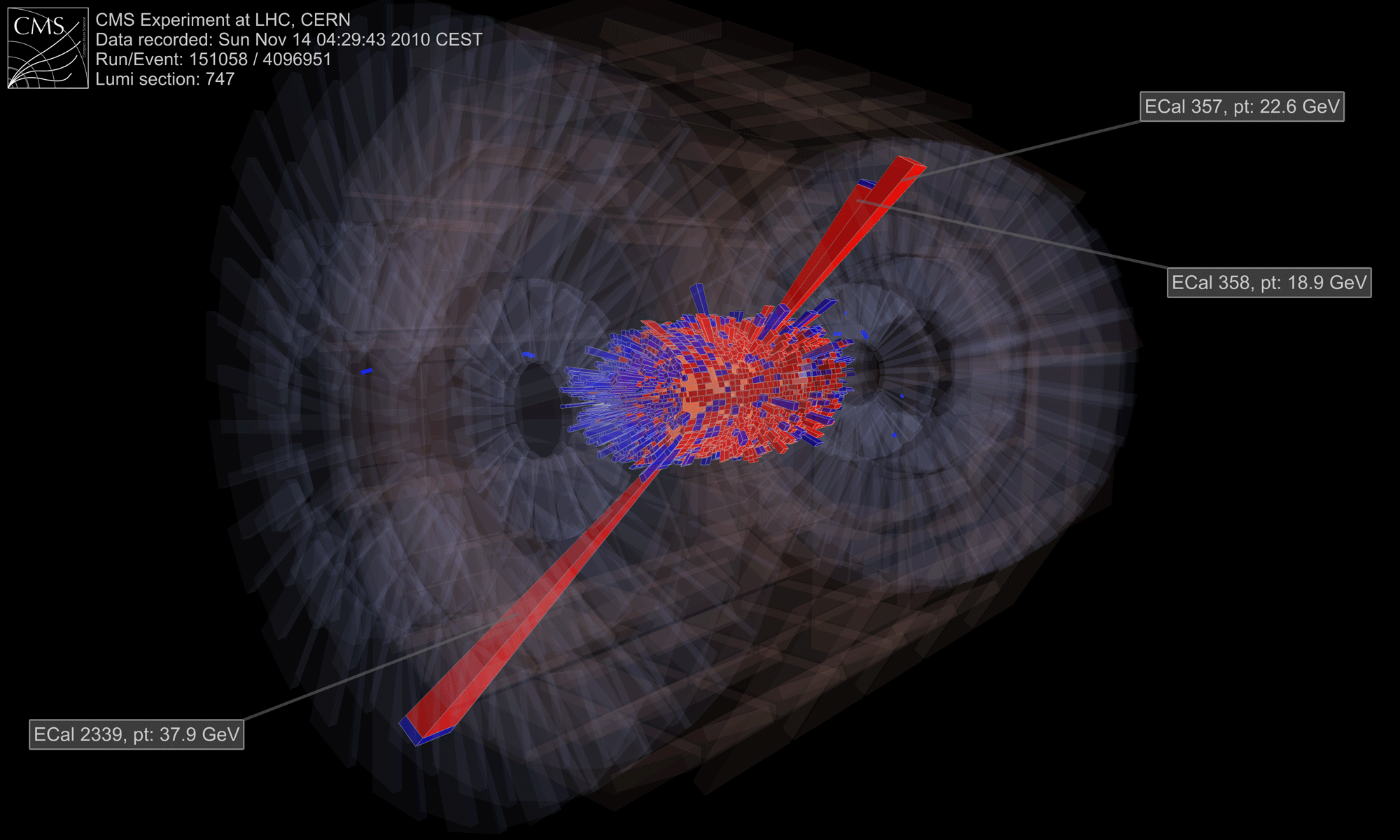}
		\caption[CMS event display]%
		{{\small CMS (Fireworks)}}
		\label{fig:ed-cms-fireworks}
	\end{subfigure}
	\caption[Examples of HEP event displays]
	{\small Different examples of HEP 3D event display applications: \subref{fig:ed-atlas}) an ATLAS 3D event display made with VP1~\cite{ATLASED072015,ATLASVP12010};
	\subref{fig:ed-cms}) a CMS 3D event display made with iSpy~\cite{CMSED}; 
	\subref{fig:ed-alice}) an ALICE 3D event display \cite{ALICEED};
	\subref{fig:ed-lhcb}) an LHCb 3D event display \cite{LHCbED}; \subref{fig:ed-belleii}) a Belle II VR event display \cite{BelleIIVRED}; 
	\subref{fig:ed-cms-fireworks}) a CMS 3D event display made with Fireworks~\cite{CMSFireworks}}
	\label{fig:screenshots}
\end{figure*}

In the following subsections several important aspects of data access, application development and distribution, and geometry description
and visualization, as they pertain to the current landscape of event displays, are discussed in more detail. Screenshots of several
event displays can be seen in Figure~\ref{fig:screenshots}.

%[ATLASED042016]
%[ATLASED072015]
%[OPEN-PHO-EXP-2015-013]
%[ALICE-EVENTDISPLAY-2017-007]
%{other images from other experiments ???}

\hypertarget{data-access}{%
\subsubsection{Data access}\label{data-access}}

Access to event data comes either natively or via intermediate formats. In the former case direct access of native event formats
is only possible for an application integrated with the experimental software framework.

There are several advantages to having access to the experiment's framework, such as full access to the experimental
data in its native format and to software tools, services, and databases. Through them, event display applications can
make use of the full detector simulation geometry, of conditions data, and of all the framework's application program interface (API).
 
One disadvantage of this approach is that full support for the display application is often limited to those platforms
on which the framework itself is supported, limiting cross-platform distribution and support.
One way to mitigate this is to distribute a light version of the framework along with the application; CMS Fireworks~\cite{CMSFireworks}
takes this approach. However, issues of platform support for the light framework and for the visualization application can still exist.
A further disadvantage to the full-framework (and even light-framework) approach is that one must also support various versions of
the data format along with the underlying framework API. In addition, users have to have knowledge of the framework in order to interactively
explore and visualize event-based data. Lastly often the user-interface to a
full-framework application is geared towards the expert.

The latter approach to data access is via an intermediate format. Usually the data needed for visualization is a subset of the
full information found in the native experimental format. Therefore one can extract what is needed from the framework through
the usage of dedicated exporting software tools and store in intermediate formats.

With the use of an intermediate data format (usually based on different flavors of the XML or JSON formats) the event display
application is potentially separate from the experimental software framework and therefore its deployment is not limited to
those platforms officially supported by the experiment. It is then possible to have both lightweight data and applications,
which can be easily and broadly distributed. %even for outreach and educational activities for the general public.

The primary drawback to this approach is that, with no direct access to the experimental data, some
information is necessarily not accessible. Moreover, every time there is the need to modify the content of the
intermediate data files, one needs to run the data extraction tools on the native data again.
In addition to that, only events which are identified as potentially interesting are extracted and their
data reduced to be stored in the intermediate data file; so, if the end user wants to analyze and visualize other
potentially interesting events, the extraction/reduction step must be performed again on the relevant data.

\begin{figure}
	\centering
	\includegraphics[width=\linewidth]{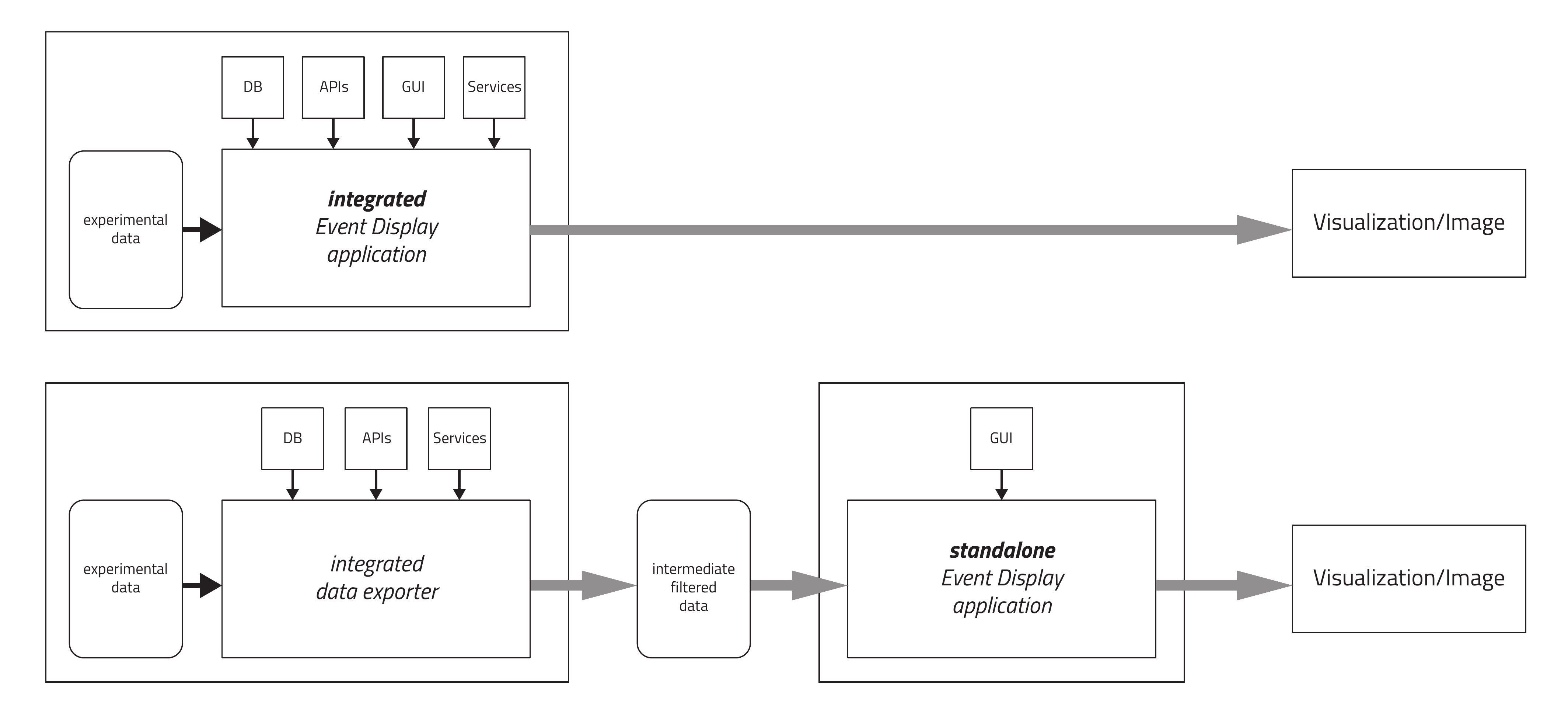}
	\caption[Different types of event display applications]{The two different types of event display applications. At the top,
	the framework-integrated event display application is able to access all experimental data, all services, APIs, and databases
	provided by the experiment's framework; as a drawback, the application must be run on specific platforms supported by the
	framework and it must use only graphics and GUI libraries compatible with them. At the bottom, the standalone approach, where
	experimental data are accessed, filtered, and extracted by using custom data exporters, which create intermediate data files
	containing only the interesting pieces of information; then, a standalone event display application reads those data in and it
	creates the required visualization; the advantage is having a cross-platform application which can use any graphics and GUI libraries,
	while the drawback is the lack of direct and full access to the experimental data and to the experiment's software tools,
	which prevents a detailed, full visualization.}
	\label{fig:eventdisplaysapplicationstypes}
\end{figure}

Regardless of the approach to data access one must consider the use-case: what is useful or necessary
in one use-case may not be for another. A graphical representation of the two approaches can be
seen in Figure~\ref{fig:eventdisplaysapplicationstypes}.

%%% MOVED TO SECTION "Suggested guidelines"
%But all these advantages come at a price, at least in the current landscape,
%as direct access to the experimental data is not possible. One has to export the required information from the experimental
%data into, for example, JSON, XML, or 3D-targeted file formats, and then render this information. However, in this process,
%one has distilled down from the experimental format and some information is lost or rather cannot be easily accessed.

%The access to experimental data is a problem in today’s experiments, and it is the most critical blocker when developing
%visualization tools for HEP. Developers have to choose from the beginning the data access pattern and the target of the
%application: either an integrated application which can visualize all data, but which can run only on specific platforms;
%or a cross-platform standalone tool, which can visualize only a subset of all the information. Moreover, the many
%differences in the access to data among the experiments, make very difficult to conceive common solutions and tools.

%Data, in fact, should be much more easily accessible to scientists, in a more transparent way and without the usage of
%complex software frameworks. That is an issue which is investigated and addressed in more details in the HSF
%"Data Organization, Management and Access" Community White Paper~\cite{HSF-CWP-2017-04}.

\hypertarget{application-development}{%
\subsubsection{Application development and distribution}\label{application-development}}

\begin{figure}
	\centering
	\includegraphics[width=0.95\linewidth]{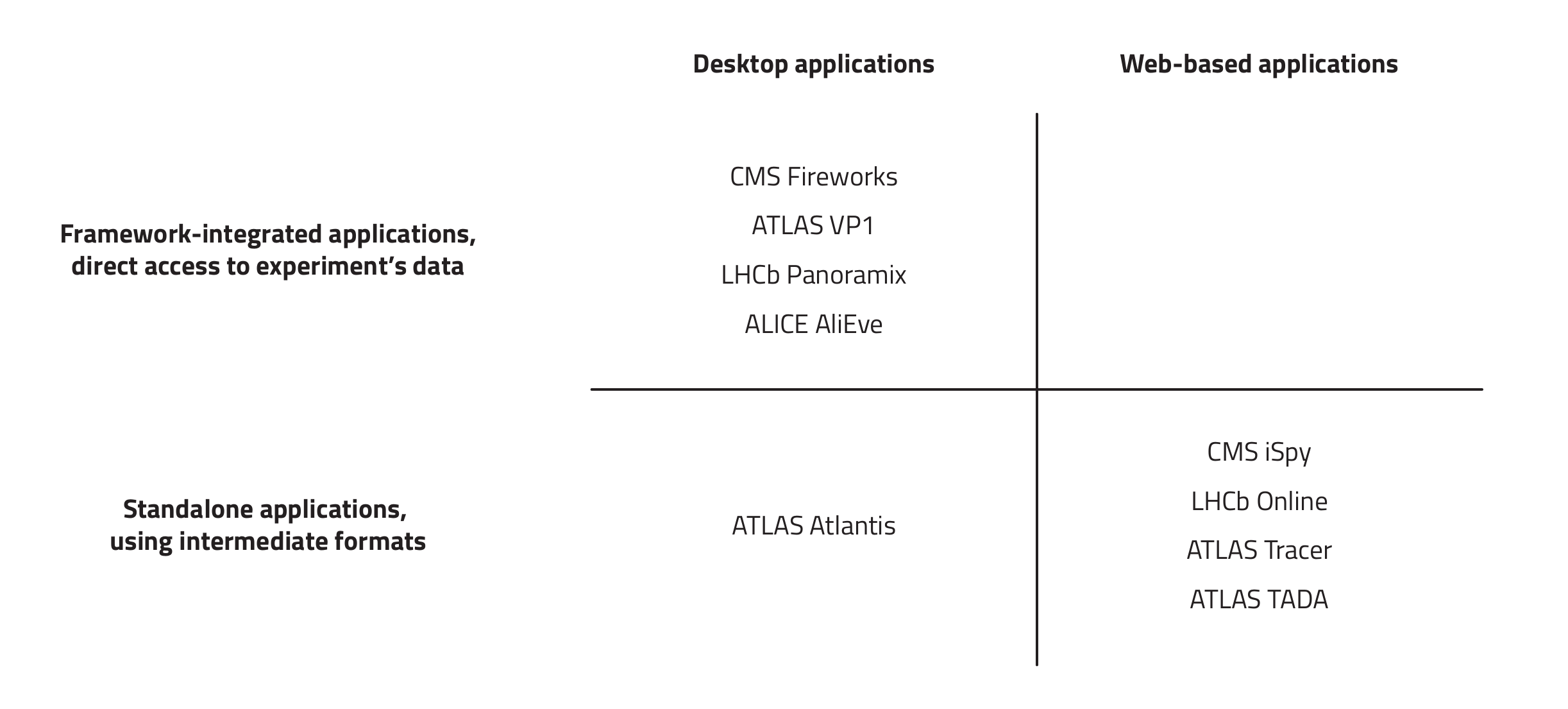}
	\caption[Current landscape of event display applications in HEP]{The image summarize the current landscape of event display applications in HEP.
	Many experiments developed full-framework desktop applications as well as light, web-based applications. As one can see from the plot,
	there are no examples, so far, of full-framework applications using web-based visualization graphics, due to the data access issues
	with today's experiments' frameworks. A new approach in that direction is what it is suggested in this paper, in Section \ref{client-server}.}
	\label{fig:currentvisualizationlandscape}
\end{figure}

Currently, the two most common ways of distributing event display applications are as a desktop application and as a web
application running in the browser. Each approach has its advantages and disadvantages which are further described in this section.
The current landscape is summarized in Figure~\ref{fig:currentvisualizationlandscape} and is further described in this section.

Native mobile applications running on devices such as smartphones and virtual reality applications are less common in HEP.
However, they are a growing feature in the current landscape and many experiments are exploring the
possibilities of those emerging technologies. At the end of this section we describe briefly the mobile applications released so far.
Further developments will be described in
Section~\ref{modern-tech}.

{\bf Desktop applications} Many experiments have developed integrated event-display applications in C++, which is the main language
used for developing HEP software frameworks, on top of the OpenGL~\cite{OpenGL1992} APIs.
The choice of the OpenGL API, compared to other APIs like Direct3D~\cite{Direct3D}, resides in its cross-platform nature as
OpenGL is an open standard. The OpenGL consortium defines the API: the interface with which all the implementations have to
comply. The actual implementation is provided by vendors, usually targeting a specific hardware. Many hardware and
software companies such as Intel and NVIDIA are part of the OpenGL consortium, which assure
the support and the lifetime of the OpenGL API.

Some HEP visualization applications use OpenGL calls directly through custom graphics engines. This is the most
robust approach as the developers can take full control over the OpenGL interface
and the project can be independent of other software libraries. Two examples of HEP applications which follow this path are the
ATLAS Persint application~\cite{ATLASPersint2012} and the ROOT EVE toolkit~\cite{ROOTEVE2007}, which is used both by the CMS
Fireworks application~\cite{CMSFireworks} and the ALICE Event Visualization Environment (AliEve)~\cite{alieve}.
One disadvantage of this approach is that developer time has to be assured to maintain the graphics engine itself,
on top of the effort needed for the development and maintenance of the event display application.

%\missingfigure[figwidth=12cm]{schematic B - OpenGL API access}

\begin{figure*}
	\centering
	\begin{subfigure}[b]{0.475\textwidth}
		\centering
		\includegraphics[width=\textwidth]{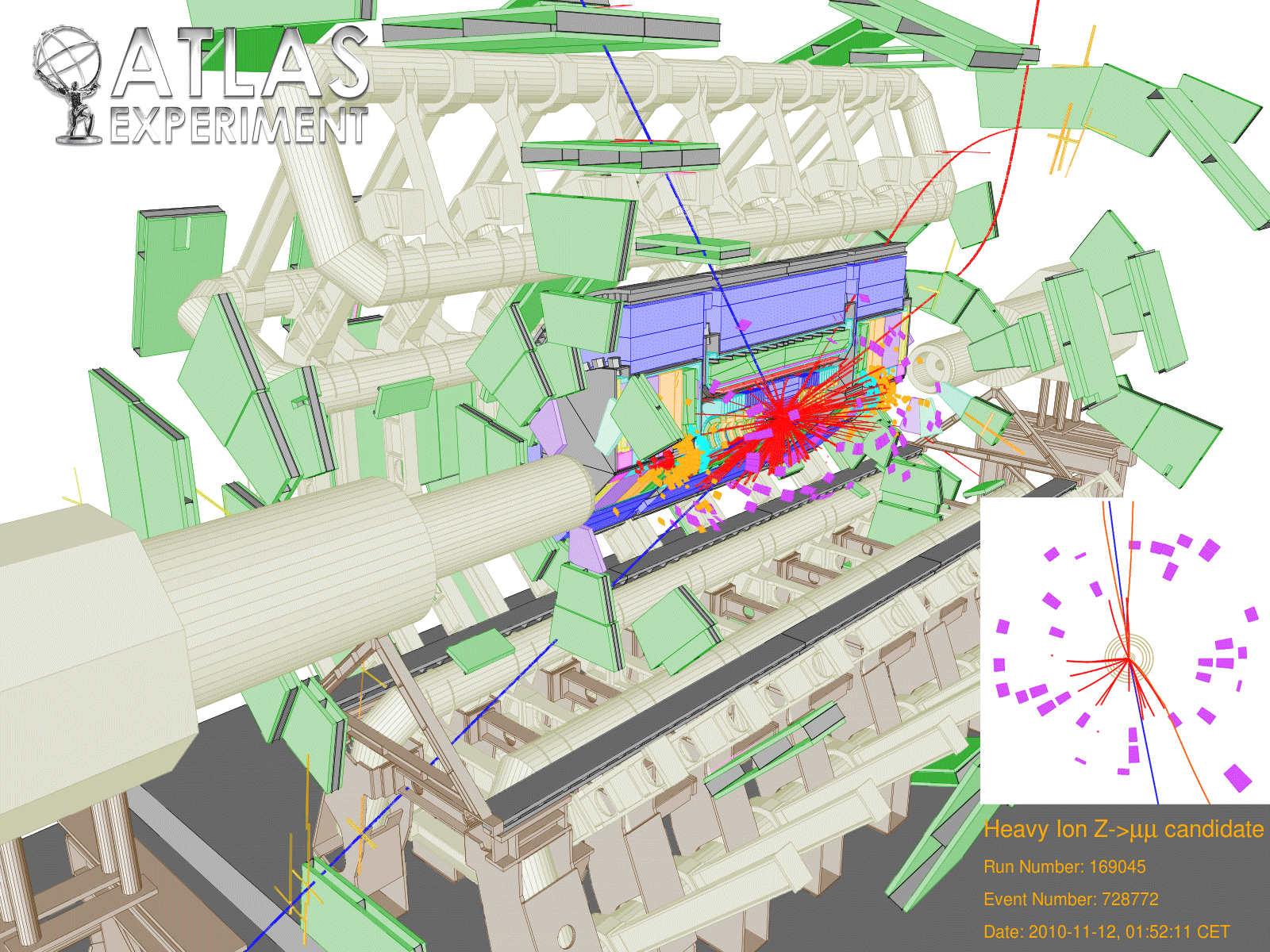}
		\caption[ATLAS 3D event display made with Persint]{{\small}}
		\label{fig:atlas-persint}
	\end{subfigure}
	\quad
	\begin{subfigure}[b]{0.475\textwidth}
		\centering
		\includegraphics[width=\textwidth]{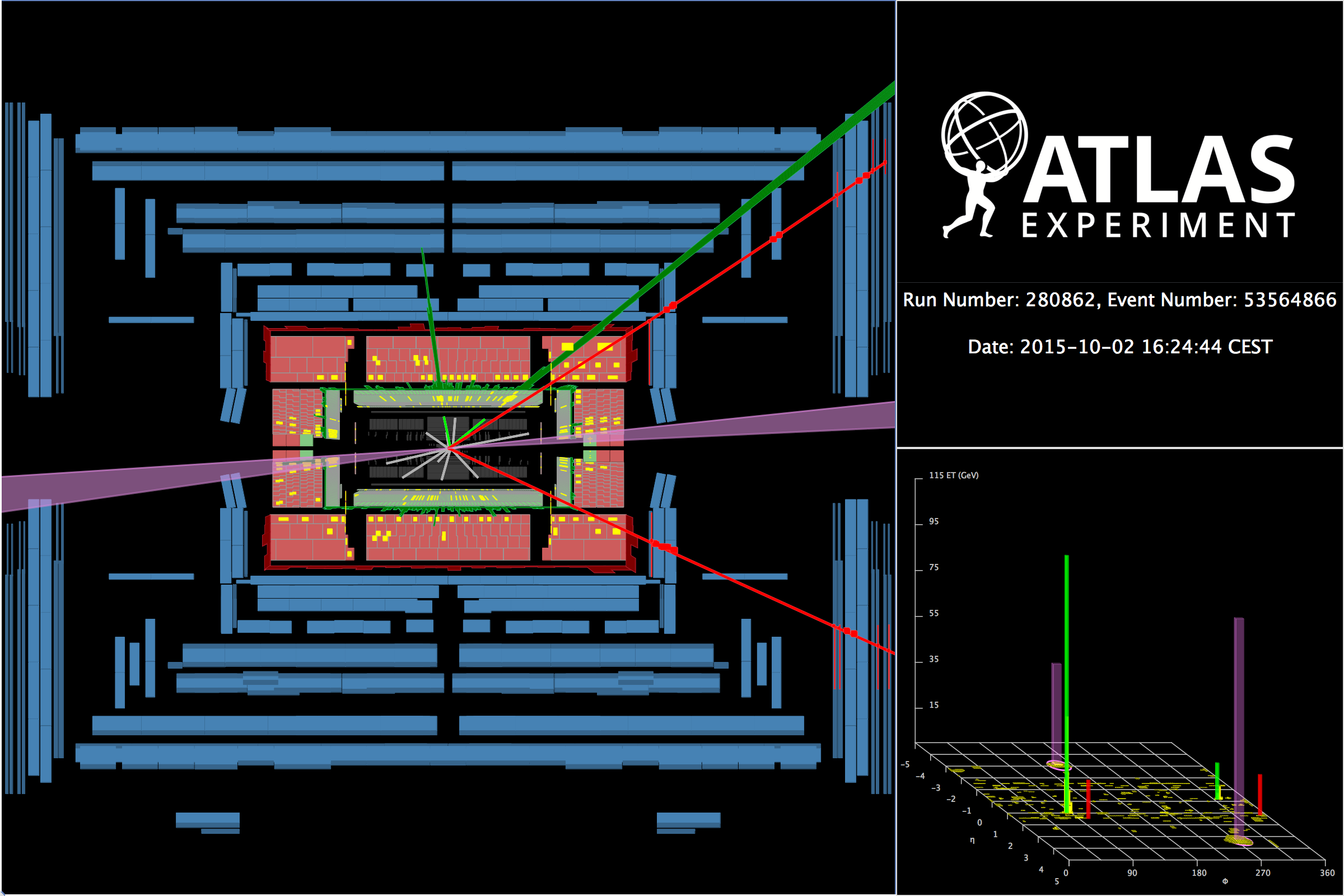}
		\caption[ATLAS 2D event display made with Atlantis]%
		{{\small }}
		\label{fig:atlas-atlantis}
	\end{subfigure}
	\caption[Using 3D editing software for HEP Geometry]
	{\small \subref{fig:atlas-persint}) an ATLAS 3D event display made with Persint~\cite{AtlasPersintZmumuED,ATLASPersint2012}; \subref{fig:atlas-atlantis}) an ATLAS 2D event display made with Atlantis~\cite{AtlasAtlantisHiggsED,ATLASAtlantis}.}
	\label{fig:cms-sketchup}
\end{figure*}

Other applications use higher-level interface libraries as graphics engines. This has the advantage of delegating
a large part of the lower-level development work to external software packages, leaving the developers to concentrate
on the application development itself. A popular graphics library used in HEP software has been
Open Inventor~\cite{OpenInventor1993}, used by the defunct CMS Iguana~\cite{CMSIguanaPaperNIM,CMSIguana} application, or
its clone implementation Coin (also known as Coin3D)~\cite{Coin3D}, which has been used by applications such
as ATLAS VP1~\cite{ATLASVP12010}, LHCb Panoramix~\cite{LHCbPanoramix} and the defunct desktop version of the
CMS iSpy application~\cite{CMSISpy}. Coin / Open Inventor was
chosen because of its integrability in C++ code, its performance, and its coding style. Moreover, the way Open Inventor handles
graphical volumes could be easily matched with the way geometry volumes are handled to describe the detectors in HEP experiments.
Open Inventor organizes geometry volumes as a series of nodes in a tree-like structure in the same way as some
HEP experiments do. ATLAS, for instance, developed their geometry library “GeoModel”~\cite{ATLASGeoModel2004} based on
the same tree-like structure of nodes used by Open Inventor.

The drawback of this approach is the dependency on external software projects, which could end up with a loss of functionality if
third-party library development and support are abandoned. Many scientific visualization applications, also in fields other than
HEP, faced this when the support of the Coin library was dropped by the company that led its development~\cite{CoinEndOfLifeLetter}.
The result is the aging of libraries which after a while show incompatibilities with modern hardware, compilers, and platforms. The time
spent by HEP developers to repair or to maintain those abandoned libraries results is time not spent on actual development
of the software applications themselves.

An additional approach to the development of event displays is to create and distribute an application using the Java programming language.
The Atlantis~\cite{ATLASAtlantis} program and its derivative MINERVA~\cite{ATLASMinerva}, which is used as an educational tool,
both developed for the ATLAS experiment, are based on the Java graphics libraries and can be run either online in a web browser
or stand-alone on a desktop machine.

{\bf Web-based applications} Several experiments at the LHC, notably CMS~\cite{CMSISpyWebGL}, LHCb~\cite{LHCbOnline2014}, and ATLAS
~\cite{ATLASTada2016, ATLASTracer2015} have created web-based event displays using WebGL (Web Graphics Library)~\cite{WebGL2011}.
WebGL is a JavaScript API that conforms to OpenGL ES (a subset of the OpenGL API for embedded systems) conceived for
rendering interactive 3D and 2D graphics within any compatible web browser without the need of external plug-ins.
With WebGL in the browser one can combine high-quality graphics with the functionality and accessibility of the browser.
This combination of graphics and user function was previously only available via bespoke desktop applications based on OpenGL and
graphical user interface toolkits such as Qt~\cite{QtFramework}. Browser-based event displays have several distinct advantages: they are
easy to distribute to the user, they can be prototyped quickly, and the client is often much lighter-weight,
as the need for building, packaging and distributing external libraries is greatly reduced.
There are also several mature and actively developed WebGL frameworks, such as the popular three.js~\cite{ThreeJS},
that provide straightforward and simplified APIs for ease of development.

{\bf Mobile applications} Mobile devices such as smart phones and tablets are more and more ubiquitous and these devices
are used more and more as substitutes for desktop and laptop machines.
However, mobile devices still do not have the computing power usually needed for HEP data analysis, where huge amount of experimental data
are retrieved and processed. In addition they usually run dedicated operating systems whose self-contained nature
makes their integration within the HEP workflow difficult, particularly for the statistical-based visualization used in
data analysis, described in Section~\ref{statistical-data-visualization}. Despite the current limitations for these use-cases mobile
devices can be found in the current landscape of event displays.

Several event display applications have been developed for, or at least can be run on, mobile devices.
An example of a native application is LHSee~\cite{LHSee} which live-streams ATLAS events, processed and extracted
through Atlantis~\cite{ATLASAtlantis},
to a user’s phone and provided contextual information on ATLAS and the
events being displayed. The CMS iSpy WebGL application~\cite{CMSISpyWebGL} runs on mobile devices in the browser.
The Camelia application~\cite{CERNCamelia}, % and its successor TEV~\cite{CERNTEV}
developed by the CERN Media Lab using the Unity game engine~\cite{Unity3D}, can be built as a mobile application
by using the tools provided by the game engine and run on mobile devices as well.
More details on game engines can be found in Section \ref{graphic-engines}.

{\bf Virtual and augmented reality applications} Virtual Reality (VR) simulates the user’s physical presence in a virtual environment,
and the application is typically run on a head-mounted display that provides visual and aural experience of the simulated environment.
Different degrees of realism and immersion are possible, depending on the targeted hardware.

VR technologies can be used to build immersive applications, to let the general public virtually visit HEP detectors and explore
experimental sites. Many HEP experiments started developing VR applications, mostly as an educational tool, for outreach events.
These include ATLASrift~\cite{ATLASRift} and Belle II VR~\cite{BelleIIVR}.
As many HEP experiments are not accessible during data taking or are classified as supervised areas due to security or safety issues,
VR applications let the HEP community open their sites and experiments to the general public. Also,
they let people look at simulated  collisions in a simplified yet realistic environment, which help people acquire the basic
concepts on which HEP experiments are built and run. Such applications are currently used in public events,
in museums and science centers, and in meetings with the governments and the funding agencies.

Augmented Reality (AR), instead, uses a camera to take a view of the real world around the user and screen where simulated
objects are rendered in 3D and shown on top of that image dynamically, following the user's interaction and motion.
This lets the user move within an environment where real and simulated objects live together. AR can be used in HEP as
an educational tool, for instance to dynamically show and describe a HEP detector to a group of people or a class.
Some HEP experiments have been started exploring AR technologies for HEP, particularly ALICE, ATLAS, and CMS.

Both VR and AR applications are usually developed in specialized graphics engines, which prevent the interaction
with the experiment's framework to access native data. More details on VR and AR applications for HEP are found in Section \ref{vr}.

%\missingfigure[figwidth=12cm]{Add ref/comment to https://twitter.com/acceleratARapp}

\hypertarget{geometry-description}{%
\subsubsection{Geometry description and visualization}\label{geometry-description}}

\begin{figure*}
	\centering
	\begin{subfigure}[b]{0.475\textwidth}
		\centering
		\includegraphics[width=\textwidth]{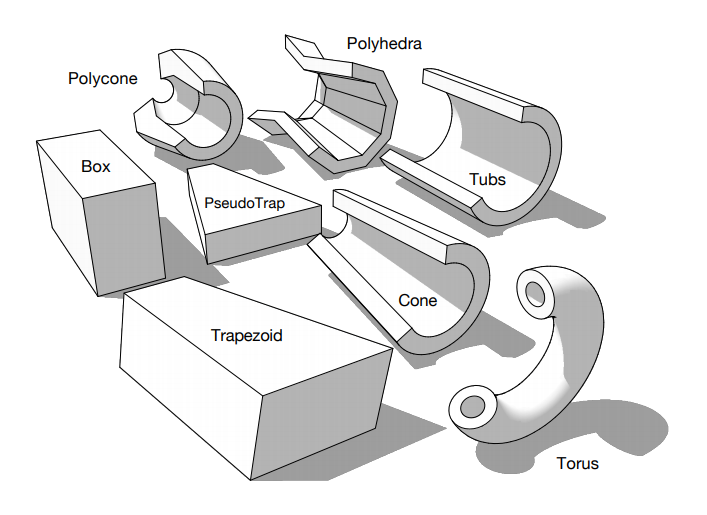}
		\caption[Converting the CMS detector geometry data to 3D meshes]{{\small}}
		\label{fig:ed-sketchup-a}
	\end{subfigure}
	\quad
	\begin{subfigure}[b]{0.475\textwidth}
		\centering
		\includegraphics[width=\textwidth]{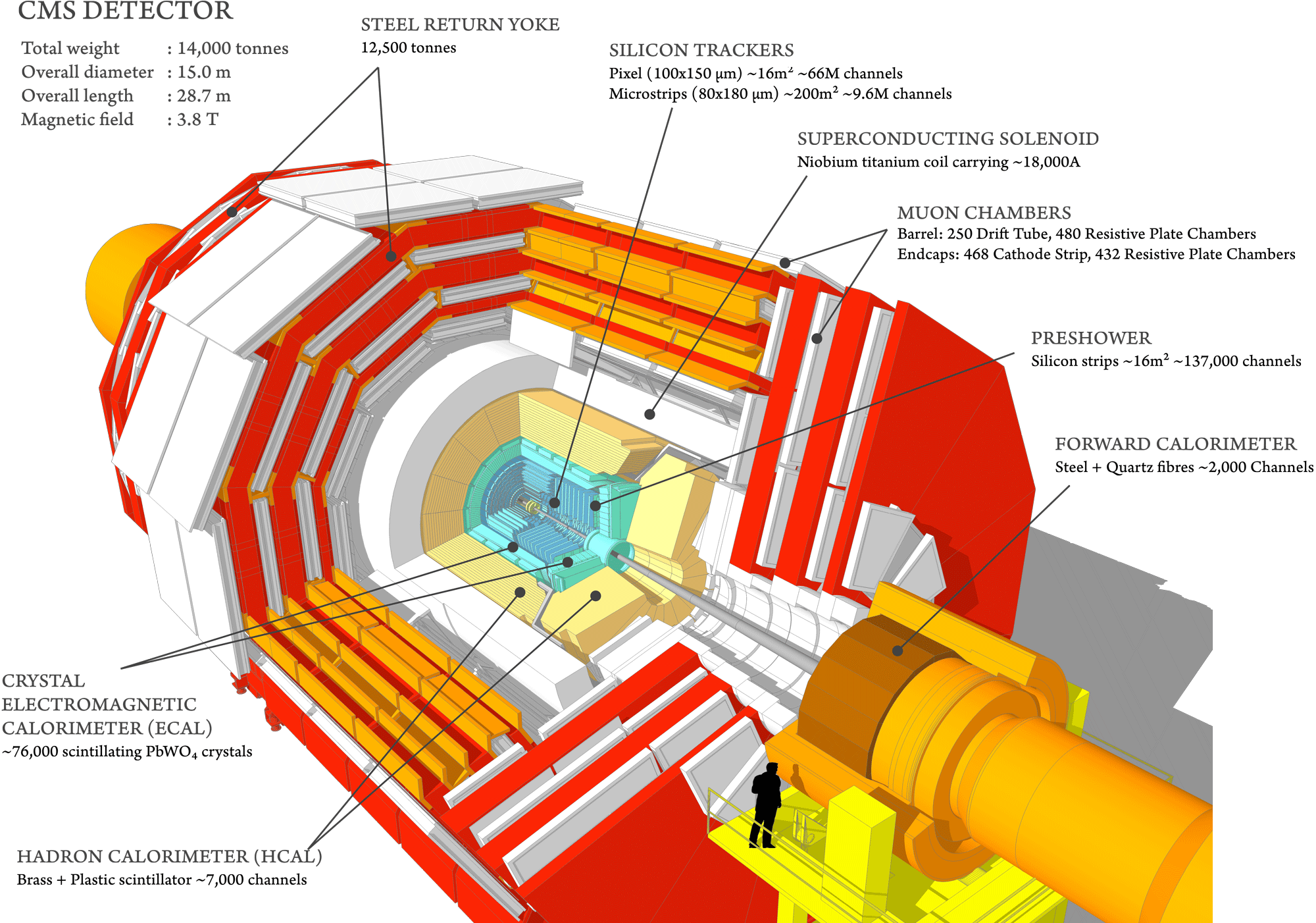}
		\caption[Rendering the CMS geometry in the SketchUp 3D software]%
		{{\small }}
		\label{fig:cms-sketchup-b}
	\end{subfigure}
	\caption[Using 3D editing software for HEP Geometry]
	{\small Use of a 3D software (SketchUp) to render the basic geometrical shapes imported from the CMS detector description and the final volumes \cite{CMSSketchUpImages}.}
	\label{fig:cms-sketchup}
\end{figure*}

Geometry visualization provides important visual context for event displays and dedicated geometry displays are useful
applications by themselves. There are typically three levels of detail found in applications. The most detailed geometry
is typically called the \textit{simulation geometry} and can include the sensitive elements of the detector as well
as support structure. Detector experts are typical end-users of applications that display this level of geometrical detail. Less
detailed is the so-called \textit{reconstruction geometry}, which describes the sensitive elements of the detector
such as calorimeter crystals and wire and strip chambers. It is this level of detail that is typically found in
event displays. The least-detailed geometry descriptions are those that are simplified versions of the
detector geometries and are used to provide visual context only.

Currently, different geometry formats and libraries are used in HEP. 
Some experiments use their own custom format (\textit{e.g.}, \cite{ATLASGeoModel2004}), while others use the geometry tools provided by the ROOT framework~\cite{Root1997}.
More recently, some attempts have been made to build common formats and libraries for detector geometry, like DD4HEP~\cite{dd4hep},
adopted in conceptual design studies for future high-energy colliders, including the CLICdp~\cite{clicdp} and FCC~\cite{fcc} collaborations.
In all cases detector volumes are built from simpler geometrical entities: geometrical shapes like Tube, Cone, Box, and
more complex variations of these are combined in order to build the volumes of the experiment’s geometry.
Geometry libraries and formats are also described in the HSF \textit{Detector Simulation} Community White Paper~\cite{HSF-CWP-2017-07}.

The differences in geometry formats used by the different experiments, by detector simulation programs like Geant4~\cite{Geant4},
and by data analysis frameworks like ROOT, typically require developers of visualization applications to write converters
between the different formats. In addition it is often not easy to use a visualization tool developed for one experiment with another one as
current visualization tools are often tightly bound to the geometry library used by the experiment.

In framework-based applications the geometry information can come directly from the experiment's detector description and in
 many cases the hierarchical structure of the detector description is preserved and accessible. Standalone applications
typically use a geometry file with information exported from the software framework. An example hybrid solution is the one developed for the CMS experiment using the  
SketchUp application~\cite{sketchupSW}, described in Ref.~\cite{CMSSketchUp} (see Figure~\ref{fig:cms-sketchup}).
The CMS detector description \cite{cmsDetectorDescription} as written in XML is parsed using Ruby scripts and 3D models are built using the
SketchUp program via its Ruby API. SketchUp can then export to various standard 3D file formats. In this way
detailed simulation geometry can be available in a standalone application.

%CMS created 3D models of the CMS detector in SketchUp~\cite{CMSSketchUp}. Ruby scripts were used to parse the
%description of the CMS detector geometry written in XML and to build 3D models in SketchUp via its Ruby API.
%Figures produced based on these 3D models have been widely used: in many conference presentations, many PhD theses,
%technical design reports, a few notable journal publications, books, magazines, posters, brochures, websites, and other media.
%Further, the 3D models were imported in iSpy WebGL mentioned above.

%In the case of framework-integrated applications, the geometry shown is usually the actual geometry used in the detector description,
%taken directly from the experiment’s software framework and used for all kinds of tasks related to geometry (see Figure [ATLASED042016]).
%While in the case of  java-based standalone or web-based applications the geometry shown is usually a simplified version of the actual one
%(see figure [ATLASED072015]) or, merely, a fake geometry.

\hypertarget{statistical-data-visualization}{%
\subsection{Statistical data visualization}\label{statistical-data-visualization}}

\epigraph{The simple graph has brought more information to the data analyst's mind than any other device.}{\textit{John Tukey \cite{Tukey1962}}}

Data visualization also means visualizing quantities and properties taken from a series of events, in order to extract statistical
meaning from them. Examples of statistical data visualizations are histograms and scatter plots.

In HEP, like most other scientific disciplines, visualization of the data and of the final results plays a key role in the
analysis pipeline. A new projection of the data may provide new insight, results must be summarized in a
clear and concise way, and multidimensional parameter spaces need to be visualized in an understandable fashion. The discussion
of how to properly display data is not new~\cite{Tufte1986}, but the tools are constantly evolving. Since its introduction 20 years ago,
ROOT~\cite{Root1997} has become the most widely used package in HEP to make plots, graphics, and even to build event displays.
It was developed at a time when there were few alternatives for the HEP community that did not have a significant
financial cost, and it has performed admirably.

However, the landscape is changing and there are several existing tools driven by non-HEP communities available. This section will look at
some of the current alternatives and comment on what options might be available in the future and what our needs are. The main focus is
on data exploration and presentation tools. The former describes tools with which one prepares and builds visualizations for the purpose
of exploring and attempting to understand one's dataset. The latter describes tools for the presentation of final plots in a convenient and accesible way.
For more details on data analysis itself, one should refer to the HSF \emph{Data Analysis and Interpretation} Community White Paper~\cite{HSF-CWP-2017-05}.

\hypertarget{stats-desktop}{%
\subsubsection{Desktop solutions}\label{stats-desktop}}

As it stands, ROOT is the most widely adopted plotting tool within the HEP and Nuclear Physics community.
It has even made some inroads to the astrophysics community and some small pockets within the financial community,
to where some physicists migrated. However, few other disciplines have adopted it. Still, it has many features beyond
the standard 1D/2D/3D histogram/graphing tools, such as 2D and 3D shapes, widgets for building a GUI, a JavaScript
implementation for web-based analysis~\cite{rootjs} and is available within a Jupyter notebook~\cite{JupyterNotebook}. But to
access the plotting features, an analyst must install the entire ROOT package which includes file I/O, scientific
libraries, fitting routines etc.,\ and often the installation process is non-trivial.

Many current HEP analysts make wide use of the Python programming language and the PyROOT libraries~\cite{PyROOT}.
Python is also very popular outside the HEP community and so it is worth looking at non-ROOT options available to Python users.
A recent (as of 2017) summary of the field was presented by Jake VanderPlas at PyCon 2017~\cite{VanderPlas2017}, a subset
of which will be presented here. It is emphasized that this is just a sampling and that the number of options available
is a function of time.

\begin{itemize}
\item The Python library \textit{matplotlib}~\cite{Hunter2007}, released in 2003, is the most mature plotting tool
for python and is the standard for most users. It can produce journal-quality graphics and there are some add-ons
that can improve the default plotting options~\cite{seaborn}. It does 1D, 2D, and 3D graphics with varying degrees
of success, but does not integrate with OpenGL libraries and so it can slow down when the number of data points gets very large.
It does produce most of the histograms found in HEP but some minimal, extra work must be done by the user
to make histograms with error bars. Plots are reactive in the sense that you can zoom in on different regions of the graph,
but you cannot do anything more significant with other mouseover commands (links, additional information, etc.).

\item The R programming language has several widely used graphics tools, both built-in or provided by external modules.
\textit{ggplot2}~\cite{Wickham2009} and \textit{lattice}~\cite{Sarkar2008} are particularly useful to visualize data in
multidimensional parameter spaces. \textit{ggplot2} is an implementation of the guidelines contained in the classic
text \textit{The Grammar of Graphics}~\cite{Wilkinson2005}, while \textit{lattice} is an implementation of the
so-called \textit{Trellis Display}~\cite{Trellis}. Both packages are very popular outside the HEP community and a wide range
of learning materials are available in books, online courses, and other media. Both packages are well developed and mature
and offer mechanisms for users to extend them by adding new features.
\textit{lattice} has a longer history: in 2005, the year \textit{ggplot2} first appeared, \textit{lattice} was already popular.
In fact, figures made with \textit{lattice} were shown in the
presentation in PHYSTAT05~\cite{phystat05} which introduced R to the
particle physics community.
\end{itemize}

\hypertarget{stats-web}{%
\subsubsection{Web-based solutions}\label{stats-web}}

Web-based data visualization is also being rapidly developed. Very sophisticated toolkits now provide tools to build web-based
fully-responsive visualization of data on all types of devices. In addition, they also offer other features, specially useful
for HEP, like full in-browser LaTeX rendering (with MathJAX) and real-time visualization of streamed data. Being JavaScript-based,
those libraries integrate with the overall ecosystem of web-based technologies, letting them use all the tools offered by other
web libraries. They are overall a good solution for data presentation, and can be combined with other tools such as Jupyter
in order to be used for data exploration. Some of the most used toolkits are described below:

\begin{itemize}
\item D3 (Data Driven Documents)~\cite{D32011} is perhaps the first web-based visualization toolkit which has been widely
adopted as the de-facto base solution for building interactive data visualization for the web. The strong point of D3 is the
link of the data to the DOM entities and the possibility to work with SVG objects natively. D3 is also the foundation layer
upon which many higher level toolkits are built.

\item Bokeh~\cite{Bokeh2014}. This is a plotting utility from Continuum~\cite{continuum}, the company behind the
\textit{Anaconda} Python distribution system and other Python modules. It is designed with the web in mind and builds in a high degree of
interactivity into the plots, making it useful to share results publicly and for building dashboards. However, it works by
writing HTML, which makes it difficult to work with unless you use specific IDEs like a Jupyter notebook. Exporting a figure
for a journal article ({\it e.g.}, in the PNG format) is non-trivial as well, as that is not currently the primary use-case for Bokeh.

\item Plotly~\cite{Plotly2015}. This is another web-oriented solution, similar to Bokeh and based on the D3 library, where plots
can be hosted in Plotly’s cloud service or viewed in a Jupyter notebook. The plots are similarly very interactive and there
are ways to export figure images, but that is not the goal. Dashboards can be built with relative ease and plotly offers
libraries in R and JavaScript, in addition to Python. They offer both a free and enterprise business model.
\end{itemize}

The so-called notebooks are a rapidly evolving way of using web-based technology for both online and offline data analysis and
visualization, with access to local resources as well. After having started from a Mathematica-like notebook user interface
paradigm mixing server-side code snippet execution, structured text, and (mostly) static visualizations, the Jupyter community
is now exploring more interactive user interface paradigms, including in the area of visualization. The JupyterLab project is
exploring a more MATLAB-like IDE user experience inside the web browser, with features such as multiple source editing tabs
and interactive python consoles. Its ipywidgets sub-project tries to make Jupyter more interactive by moving more visualization
work to client-side Javascript and introducing classic GUI widgets such as sliders and checkboxes for interacting with the live visualization.
The Belle II experiment has started to use Jupyter notebooks to train new users in data analysis; the learning curve is much gentler than in
traditional terminal-based tutorials and the time to useful visualization of results is much faster.

\hypertarget{stats-issues}{%
\subsubsection{Issues}\label{stats-issues}}

{\bf Separating data visualization from the data analysis:}
The suite of statistical plotting tools in ROOT, Matplotlib, etc.\ are adequate for analysis, and their development is
very responsive to analysts’ needs. However, it is often hard to separate plot-making abilities from the data analysis framework.
As a consequence, if a physicist’s data can only be found on a particular server, the plot-generating code must also be located
there and the outcome is sometimes hard to bring to the physicist’s laptop screen. In the worst cases, graphics files (PNGs) must be copied
from the server to the laptop for viewing. This causes a high interaction latency, discouraging exploration. That’s why the
development of new tools should go towards a sharper separation between the computation on data and the interactive data visualization
routines, pushing the latter to the client side as much as possible.

{\bf Separating the plotting functions and content from the plotting style:}
Another symptom of the tight coupling between data analysis infrastructure and plotting is that trivial changes to the plot— axis labels,
colors, and such— are so deeply buried in the analysis script that persistifying changes to them often requires a full recalculation
of the statistics. Changes to the final plot through the usage of on-display user interfaces, in fact, are overwritten and lost if a
plot is updated for other reasons (new version of data upstream, for example). Here, one could use inspiration from the increasing
separation of logic and presentation that is occurring in GUI toolkits (see {\it e.g.}\ use of CSS stylesheets in the GTK/GNOME environment).
%\end{itemize}

A looser coupling between style and content, as well as a looser coupling between locality of computation and locality of rendering,
would benefit the physics community.

\hypertarget{non-spatial-visualization}{%
\subsection{Non-spatial visualization}\label{non-spatial-visualization}}

\begin{figure*}
	\centering
	\begin{subfigure}[b]{0.6\textwidth}
		\centering
		\includegraphics[width=\textwidth]{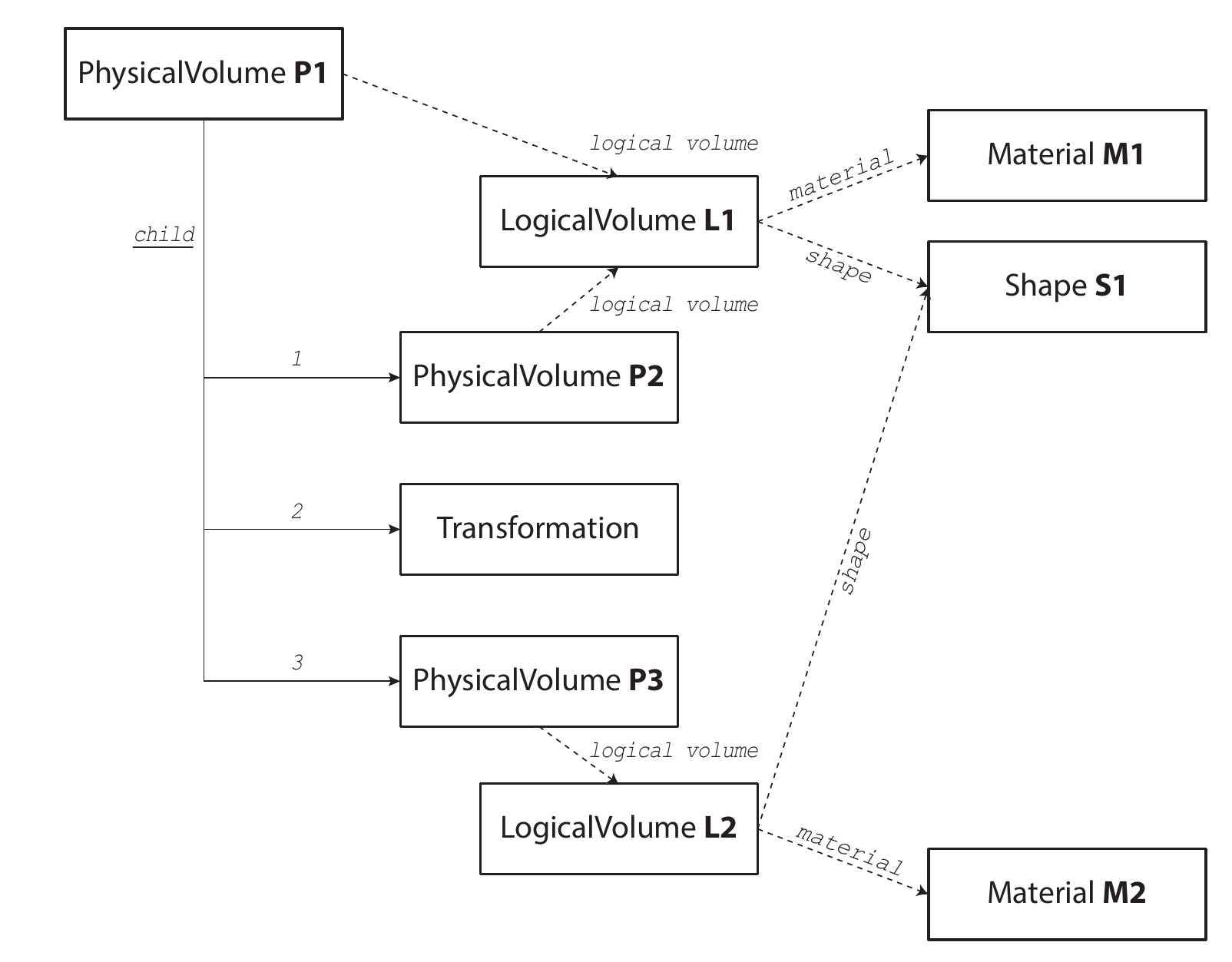}
		\caption[Tree-structured geometry data, an example from ATLAS]%
		{{\small {}}}
		\label{fig:atlasgeo1}
	\end{subfigure}
	\vskip\baselineskip
	\begin{subfigure}[b]{0.6\textwidth}
		\centering
		\includegraphics[width=\textwidth]{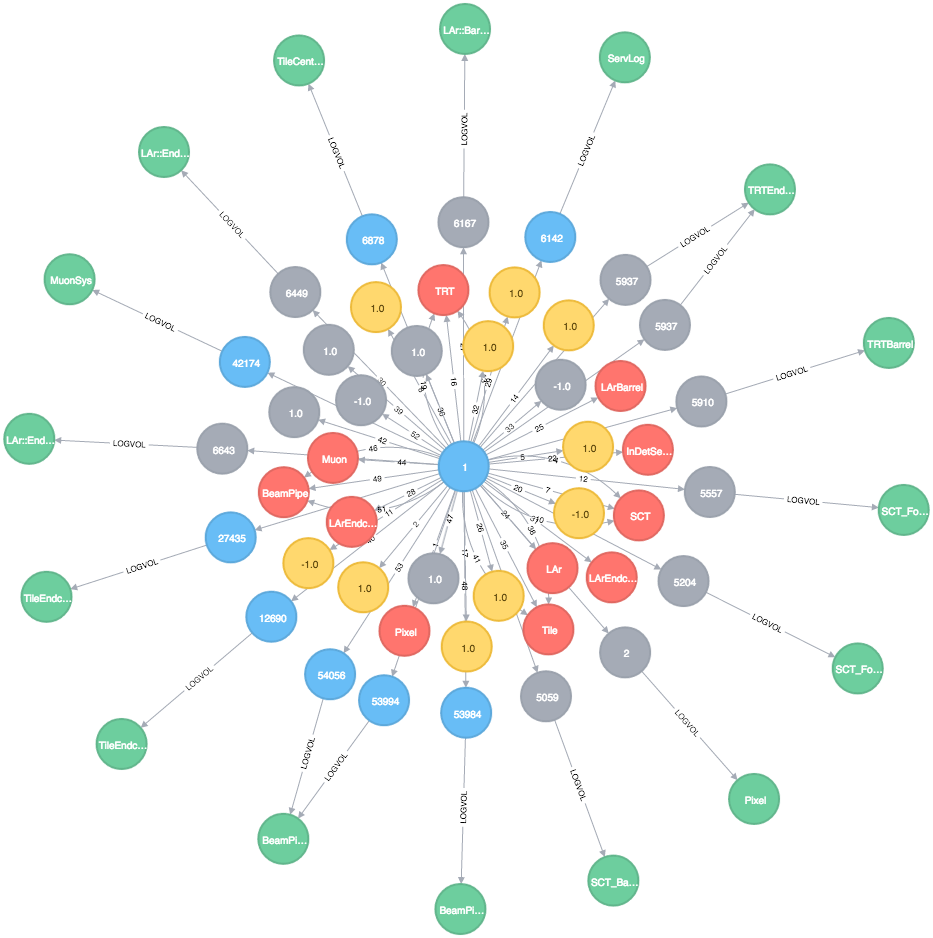}
		\caption[A graph visualizing the ATLAS geometry data]{}
		\label{fig:atlasgeo2}
	\end{subfigure}
	\caption[Using 3D editing software for HEP Geometry]
	{\small \subref{fig:atlasgeo1}) A schematic drawing depicting the tree-structure of the data describing the geometry of the ATLAS detector. Such data structures are better visualized using graphs and networks. \subref{fig:atlasgeo2}) A graph visualizing the first layer of the nodes of the ATLAS Detector Description. Different colors indicate different types of nodes; also, the labels along the lines state the different types of relationship between two data nodes \cite{ATLASGeoModel2017}.}
	\label{fig:atlasgeotree}
\end{figure*}

In HEP, there are data which are organized in a tree-like structure, and for which a graph or a network visualization is the best choice.
The Detector Description is an example of a source of such data: it describes all the pieces which compose a HEP experiment detector.
The different pieces of the Detector Description are interconnected through different relationships: geometrical volumes can be organized
in a parent-child relationship, or a property node can be shared among many volumes. The visualization of those data in a network helps
developers in the understanding and the debugging of the Detector Description, by visualizing the relationships among all the nodes and
their properties. An example, from the ATLAS experiment \cite{ATLASGeoModel2017}, of a graph visualizing the inner
structure of a HEP detector description can be seen in Figure~\ref{fig:atlasgeotree}. 

Networks and graphs are very effective ways of visualizing tree-like data, because they are able to show all the 	nodes, their relationships and their properties in a proper way. Some degree of interactivity can let the scientists applying different filters and layout, helping them to get rid of the clutter, to better understand and analyze the data.

Another example of HEP data that can benefit from a graph-based visualization is that one describing the execution chain of the jobs
used to filter and reconstruct the experimental data. Very recently HEP experiments began to develop new parallel frameworks
to concurrently handle analysis or reconstruction jobs, to efficiently exploit the parallelism offered by the modern hardware
(more details can be found in the HSF \textit{Event/Data Processing Frameworks} Community White Paper~\cite{HSF-CWP-2017-08}).
The jobs are handled by a scheduler, which organizes them according to their needed input and output data. The outcome of the scheduler
is a directed acyclic graph (DAG). The visualization by means of a graph helps the developer understanding and debugging the
reconstruction and framework code.

\begin{figure*}
	\centering
	\includegraphics[width=0.95\linewidth]{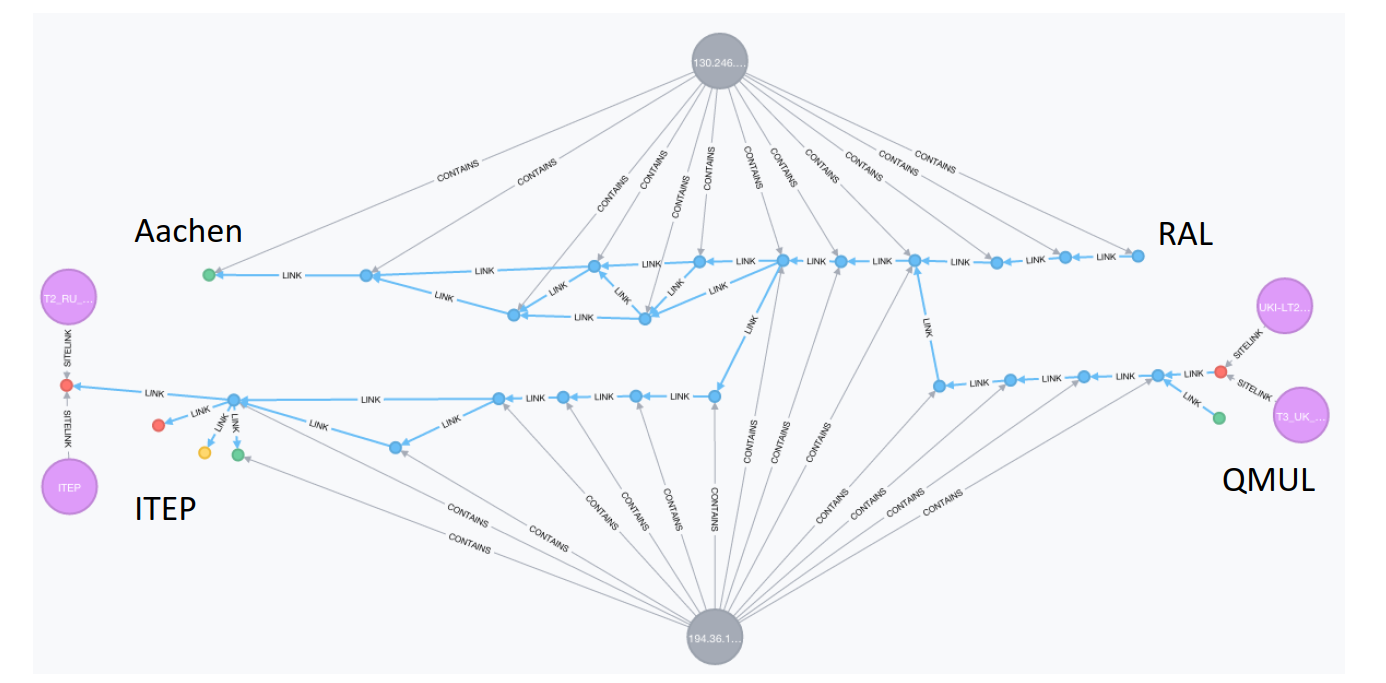}
	\caption[Graphs used to visualize network topologies]
	{\small The image shows a graph used to visualize particular snapshot of the network topology between four WLCG sites, which was used to perform network path analysis. [Image provided by the authors of \cite{neo4jnetworkstudy}].}
	\label{fig:networkstudy}
\end{figure*}

Other HEP experiments use graph-based tools to analyze and visualize other types of data, like geographical distribution and load of computer
networks used to transfer data between GRID sites \cite{neo4jnetworkstudy} (see Picture \ref{fig:networkstudy})
 or to store and query conditions data \cite{Clemencic:2012cw}.

All those data are not space- nor time-dependent, and they are better visualized through a graph or a network. Graph-based visualization,
as well as graph-databases, are fairly new in the HEP landscape but they can be very powerful tools to effectively visualize
non-spatial data which are by their nature organized with a network layout. We suggest the community further explores those tools,
to better understand the possibility offered by graph-based solutions for HEP needs.

% More examples from the other experiments, if any and if possible... ==> I didn't find any other example]

\hypertarget{suggested-guidelines}{%
\section{Suggested guidelines and future development}\label{suggested-guidelines}}

As a community what we want to suggest here is the design and the usage of common base visualization guidelines,
to be able to share knowledge and best practices among the different groups, and to foster collaboration among the HEP experiments.

\hypertarget{common-format}{%
\subsection{A common community-defined format}\label{common-format}}

Visualization has a key role in the lifecycle of a HEP experiment, addressing input data from many sources and in many different formats.
The input data are often quite tightly bound to a specific experiment's software framework and because of that many visualization tools
are integrated into the frameworks to some extent. However, the visualization is often the last step on the experiment data chain and the
output of a visualization application is often not used by any other tool in the software framework. So while the direct interaction with experimental data
formats is highly experiment specific, there is a real possibility of having the final stages of the visualization pipeline
shared between several experiments.

Let us take the example of the detector geometry. There are very many different geometry libraries and formats in use among the HEP experiments.
However, geometry libraries are all different ways to describe and handle basic geometrical entities and combinations of them.
From a visualization point of view, the output of all geometry libraries are mere descriptions of 3D shapes, which could be abstracted
from the underlying actual implementation. A shape like a box, or a cone, or a tube, or some boolean combination of them, could be
interpreted and handled the same way by a visualization tool in all experiments.

The same reasoning made for the geometry can be done for the event data. Of course different experiments detect different objects
and measure different quantities, but there are many common entities, especially among experiments within the same research field.
For example, all experiments working on hadron colliders use the notion of particle track, which is usually constructed translating
the track measurements into space points or points and angles, and use the notion of particle jet, usually
visualized as a cone whose length is related to its energy and whose radius is linked to the specific algorithm used for the jet
reconstruction. Both of these objects are currently often handled and visualized differently in different experiments, but could in principle
be the target of a common definition within the community. If so, experiments could share best practices or snippets of code,
if not complete basic tools, to build and handle their visualization. In this way, the know-how and the tools linked to visualization needs
could be shared as well, and developed in a collaborative way.

%Visualization applications should not be aware,
%and should not care, of how the given 3D volume has been created in the experiment’s code; they should only be able to
%correctly interpret the final information about it and convert that to pixel values, to be displayed on the screen.

%Of course, the usage of a common geometry library would solve many portability issues. But we know that this is not a viable solution: experiments
%have very different needs, in terms of coding languages, available know-how, integration with other software, portability, and so forth;
%and so they have to choose the geometry library which can fit the most of their needs.
%Moreover, the experiments can decide to extend the geometry library they use, in order to add custom shapes or specific functionalities.
%So, a common geometry library for all HEP needs cannot work. But we could imagine common guidelines and, perhaps, tools to visualize final geometry volumes.
%If we can imagine of having basic functions to visualize a community-defined Box shape, for example, we could imagine sharing the knowledge and
%the workload among different experiments and communities. This does not mean that all experiments have to use a common definition
%of a Box shape; but all experiments, whatever definition of a Box shape they use internally in their code for other reasons, should be able to
%provide exporters from their internal definition to the community-agreed definition of a Box shape.

We as a community still need to propose and design such common definitions and guidelines. This will
be addressed in the second phase of this Community action, following the completion of the Community White Paper.

For the moment, we observe that several event displays in experiments have exporters to translate geometry information to standard
formats used in the communities outside HEP, mainly in computer graphics and engineering. For example the Belle II experiment has
written exporters from Geant4 data to different formats, including VRML~\cite{vrml} and FBX \cite{BelleIIGeoExporters}, which are two of the most
common formats used to store and share 3D graphics data. The Unity game-development engine~\cite{Unity3D}, in turn, can export the
FBX geometry to the glTF format~\cite{glTF}, an emerging royalty-free specification for 3D objects and scenes, for fast web distribution
via the SketchFab community repository~\cite{SketchFabBelleII,SketchFab}. Displays based on three.js have access to multiple importers and exporters
of several geometry formats, including the ones mentioned above. %For the near future, we will foster similar experiments within the community.

In order to start sharing knowledge and to start working on demonstrators to show and share best practices, we propose to start
defining a common format to exchange data among the experiments. We should start by finding and listing common shared objects from
geometry and event data. After that, we should start converging on a shared definition of those objects, to build a common design
toward a data model to handle and serve them. The idea, in fact, is to enable usage of this common format to visualize data from
the different experiments with the same shared best practices, if not the same foundation software tools.

We think that community-developed common formats and tools should also be extendable, to let the experiments add their own custom
content and objects. As an example, calorimeter cells can be of very different shapes, and an experiment might need to add its own
custom shapes to the common format to visualize them properly. Thus, in addition to the part handling the common objects, there should
be a part of the format targeted at storing extended custom content, specific to a given experiment. For such custom content,
experiments will have to develop custom visualization tools as well; but they could build them upon the foundation
of the community-driven part.

Some experiments in that direction have been performed within the community in the already. For example, the
ALICE experiment made use of the mini ``Visualization Summary Data'' (VSD) set of classes, contained in the ROOT Event
Visualization Environment (EVE), to make ALICE data visualization decoupled from the AliROOT experiment’s framework \cite{TadelALICE}.

\hypertarget{serving-data}{%
\subsection{Serving the geometry and event data through services}\label{serving-data}}

Once a common format for shared objects is defined, we believe that the design and the development of online services to query and serve the geometry
data would be a very useful addition to the landscape. %, compared to what the HEP experiments currently feature.
The main driving force is the realization that detector description should be much more accessible than it is today.
For many experiments, accessing the detector description means starting and running at least parts of the experiment’s framework. The need of accessing a specific geometry version, in fact, is critical for reconstruction and simulation. However, the geometry
data needed for event visualization can often be simpler: even when showing the actual geometry of the experiment, accessing the latest
alignment constants is not crucial for visualization purposes, because small differences in the geometry are rarely visible in an event display.
So, we think that serving a “frozen” version of the experiment’s geometry would be enough, and that a simpler way to retrieve it
should be designed, to ease data access for visualization: for example, an online service could remotely serve the geometry data to the visualization applications in a standardized format.

It would be desirable for the new mechanism to have a search/filter functionality too, to let client applications query for a specific
subset of information and for a way to select the level of details, to set the desired accuracy and complexity of the retrieved geometry.

The same could be envisaged for event data, even though that is a more complicated task, involving many different layers and services:
very often event data are stored on the Grid and very often they need to be processed in order to be usable for event displays.
However, experimental data should be made more accessible for visualization. Thus, in collaboration with
the HSF \textit{Data Access and Management} working group, an API or a service to get streamed event data will be designed.
In addition, simulation data description for visualization could be handled in the same way, by implementing converters from
generators and simulation applications.

After a first phase of development and stand-alone testing, the streamed data could be used by the current visualization tools as well,
as the first step of their modernization and towards the usage of common community-developed techniques and best practices.

\hypertarget{client-server}{%
\subsection{Client-server architecture for geometry and event data visualization}\label{client-server}}

After common data formats and mechanisms to serve them are designed, together with a set of exporters required to translate the experiments’
data to the common format, we are proposing to build a client-server architecture, upon which next generation visualization applications can be built.

The idea behind that is that if we can send commands from the client to the server, and get the answer back in the data stream, then
we will be able to interact with the experiment’s framework as well, in addition to using common visualization applications to visualize
the common objects. In this way, we can develop a modular architecture where HEP experiments can share the design, the
development and the maintenance of common visualization tools, while maintaining a certain degree of freedom to add custom content
and objects and to interact with their own framework to retrieve specific content.

\hypertarget{modern-tech}{%
\subsection{Exploring modern technologies}\label{modern-tech}}

\hypertarget{graphic-engines}{%
\subsubsection{Graphics and game engines}\label{graphic-engines}}

So far direct OpenGL, its derivatives (like WebGL), or old graphics libraries have been used in HEP visualization applications. Nowadays,
another type of graphics library is rapidly evolving, those embedded in so-called game engines. Game engines are software frameworks targeted
at the gaming industry, and they feature very efficient, optimized, and modern 3D graphics.
Integration with existing code is not easy, however, because they are usually meant to be used as development environments,
and not as embedded libraries like those commonly used in HEP. So they would probably require some major changes in the usual
software architecture used in HEP. But they offer very optimized graphics and modern features, like tools for Virtual Reality,
which could be exploited in our applications.

Some HEP experiments have recently started to successfully use them to build visualization applications and event displays,
like Belle II~\cite{BelleIIVR} and the Total Event Visualiser (TEV) of the CERN Media Lab~\cite{CERNTEV},
which used the Unity game engine~\cite{Unity3D}, and ATLAS which used the Unreal Engine~\cite{EpicUnreal} for
its virtual reality application ATLASrift~\cite{ATLASRift}. These two game engines are the most popular ones on the market, and they are
free for educational and non-commercial projects.

Unreal Engine is fully open source and it supports two modes of development (C++ and the so-called Blueprints~\cite{UnrealEngineBlueprints}) that can be used
interchangeably even in the same project. It produces extremely performant executables for basically all platforms
(Windows, OSX, Linux, iOS, Android, Web, all VR platforms). All parts of development cycle are fast even for a novice,
thanks to powerful tools implemented as plugins. They have large developer communities and are very fast in supporting the latest technologies;
for example it already supports Vulkan~\cite{vulkanAPI}, the cross-platform 3D and computing API.

The Unity development platform~\cite{Unity3D} is very intuitive for novices as
well as experts and provides rapid turnaround during the development cycle: the project can be executed immediately without having to
compile and link an executable. All of the platforms supported by Unreal are supported as targets by Unity as well. Presently, user code is written in either
C\# or an adaptation of Javascript.

%These languages are somewhat foreign to the majority of HEP-trained programmers who are more comfortable with
%C++; however, this accounts in part for the broader use of Unity than Unreal.

%and VR-animation performance is reported to be significantly better on low-end devices.
%(Notably, the emerging WebVR standard~\cite{WebVR} (recently replaced by WebXR \cite{WebXR}) is not yet supported for ``VR in the browser''. It is possible in Unity to create a
%WebGL~\cite{WebGL2011} executable of a VR app; however, no browser will run this app successfully because of its need to load the
%VR-hardware’s object library, which violates security restrictions in the browser.)

Another game engine that has gained attention in recent years is the open-source engine Godot~\cite{Godot}.
While it is still not quite at the same level as Unity or Unreal Engine, it allows the deployment to similar platforms as Unity and Unreal
Engine but is very lightweight.
It offers support for 2D and 3D graphics and for multiple programming languages {\it e.g.}\ GDScript (a Python-like scripting language),
 C\# 7.0 (by using Mono), and C++.
It also offers visual scripting using blocks and connections and support for additional languages with
community-provided support for Python, Nim~\cite{nimLang}, D and other languages.

As a community, we would like to explore further the features those modern game engines can offer. Also, we would like to take a
look at possible usage patterns in the context and within the workflow of HEP visualization.

Finally, another new entry in the 3D graphics engines landscape is Qt3D~\cite{Qt3d}. The key feature of Qt3D is that it is natively integrated with the Qt framework, which eliminates a layer which was needed until now, \textit{i.e.}, a glue
package to connect the Qt GUI with the window showing the 3D content (like, for example, the SoQt~\cite{soqt} package used by ATLAS). By eliminating that, we could simplify the architecture
of our visualization tools and lower the maintenance workload. Qt3D is still in development, but it shows an initial set of
features which are worth a further consideration of the new toolkit. We plan to take a look at its development in the near future,
to see if it can satisfy the HEP requirements. Also, being open source, we could consider contributing to the Qt3D software
project as a community, by providing the pieces we need for our applications.

\hypertarget{web-based}{%
\subsubsection{Web-based applications}\label{web-based}}

Web-based graphics have traditionally been considered not powerful enough to handle the thousands of volumes that can be
shown in HEP event displays, for example, when visualizing hits in a very busy event. But the technology has rapidly
evolved and web-based graphics can now visualize very complex scenes.
For example, the glTF~\cite{glTF} 3D model of the Belle II detector~\cite{BelleII}, with tens of thousands of elements,
can be loaded, viewed and manipulated in a web browser, even on a smartphone, very effectively~\cite{SketchFabBelleII}.

The advantages of visualization in the browser have been mentioned
previously and several application have already been developed.
There is therefore already a strong interest in the community in supporting the usage and the development of web-based tools.
In particular, JSROOT~\cite{rootjs} could be used as an
underlying layer for event data visualization as well as three.js~\cite{ThreeJS}, which have been
used successfully by different experiments (\textit{e.g.}, \cite{ATLASTada2016,ATLASTracer2015,CMSISpyWebGL}) to visualize geometry and event data.

\hypertarget{vr}{%
\subsubsection{Virtual and augmented reality}\label{vr}}

As briefly described in Section \ref{application-development}, Virtual Reality (VR) describes the simulation of the user’s
physical presence in a virtual environment. This simulation
is typically delivered via a Head Mounted Display (HMD) that provides visual and aural experience of the simulated world.
Rotational and positional tracking of the user’s head and hand motion (when available) allow for interaction and navigation
in the virtual environment. This lets the user live an immersive experience of the virtual world offering many degrees of freedom. Purely rotational motion is usually refered to as
3 degree-of-freedom motion; when combined with positional motion support, an application is said to support 6 degrees-of-freedom.

There are several ways to deliver VR to the user with varying levels of functionality, accessibility, and cost.
They range from applications running on a mobile phone viewed through simple headsets to the most realistic and
immersive VR experiences provided by the combination of advanced HMDs and desktop computers.

The simplest and most inexpensive way to deliver VR is via the web browser on a mobile phone and viewed through a
Google Cardboard~\cite{GoogleCardboard} headset (which can itself be literally made from cardboard).
Rotational tracking is achieved through device orientation
controls, either in a native application or using the HTML5 device orientation API in the browser. No hand controller is
used in the Cardboard but for native applications a click event is available via a magnet attached to the Cardboard viewer.

The Google Daydream~\cite{GoogleDaydream} is the next iteration of viewers developed by Google for their Google VR technology.
Content is still delivered by a mobile phone (thus, the performance is limited by the computing power of the phone) but a
Bluetooth-connected hand controller with rotational tracking is available. The Samsung Gear VR platform~\cite{SamsungGearVR}
is another headset powered by an inserted smartphone targeted for the Oculus platform.

Currently, the most immersive and interactive VR environments are provided by the Oculus Rift~\cite{OculusRift} and the
HTC Vive~\cite{HTCVive}, which combine the computing resources of a desktop machine with sensors, controllers, and high-quality HMDs.
Those sophisticated devices are relatively expensive (even if prices are lowering in the last months) and require a fairly powerful computer to run.
The presence of a proper computer solves the performance issue of phone-based viewers, since the 3D graphics computations are handled
by the computer, but that also increases the required initial budget for people willing to test such technologies, and that limits the number of
people getting access to such platforms, for example in public events organized by HEP institutes.

Meanwhile, a new type of device has been recently developed: a standalone viewer, equipped with an on-board CPU, able to run
medium computation-intensive applications. Those devices lower the budget needed to develop and deploy VR applications significantly.
Examples of those new devices are Oculus Go\cite{OculusGo} and the stand-alone Lenovo headset for the
Daydream environment~\cite{LenovoMirageSolo}.

Game engines, described in Section~\ref{graphic-engines},  provide powerful integrated development environments for
creation of VR applications targeting multiple devices. Thanks to engines' abstraction of third party VR libraries,
most HEP experiments should be able to develop VR applications which natively support both standard displays and all VR hardware.
Currently, both ATLASrift~\cite{ATLASRift} and Belle II VR~\cite{BelleIIVR} support Oculus, HTC Vive,
and standard 2D displays. CMS, using the Unity game engine, is currently working on CMS.VR (to be released), targeting Oculus and HTC.

Augmented Reality (AR) applications use the device's camera to looks at the world around the user, to use that as an underlying layer,
over which, based on the user's interaction and motion, they dynamically render simulated virtual entities. The user can navigate in the real world, while
looking at and interacting with the virtual objects. AR technology can be used in HEP as an educational tool, for instance to
dynamically show and describe a HEP detector to a group of people or a class.

ATLAS and ALICE researchers have started to explore the possibilities offered by augmented reality for outreach and education,
by using Unity~\cite{Unity3D} and the Vuforia framework~\cite{VuforiaAR} (commercial, but free for development). The ATLAS-in-Your-Pocket~\cite{AtlasPocket} application
uses printed marks to place a rendered geometry of the ATLAS detector on top of a view of the real world in front of the user.
The More-Than-ALICE~\cite{MoreThanALICE} application allows users to superimpose a description of the detector or event visualizations to the camera image of the actual
ALICE detector (for example, on a screen or during public visits to the experimental site) or its paper model.

Development of web-based VR and AR applications for mobile browser can be done using a WebGL library such as three.js~\cite{ThreeJS} and using the
HTML device orientation control API. The viewport is split into two views, one for each eye, each with a dedicated camera view separated
by an appropriate distance to create a stereoscopic effect ({\it e.g.}\ iSpy WebGL~\cite{CMSISpyWebGL} has a stereo mode for Google Cardboard). 
The developing WebVR specification~\cite{WebVR} provides interfaces to VR hardware via the browser. A powerful framework for
development of VR applications for various devices using the browser is A-Frame~\cite{AFrame}. A-Frame has support for several device controllers as well,
such as for the Daydream and Oculus. CMS is exploring the possibilites of A-Frame in the
browser for both VR and AR ({\it e.g.}\ CMS A-Frame prototype~\cite{CMSAFrame}).

VR and AR applications could be used, in principle, to build event displays and data visualization tools for research work as well.
However, the current data access model prevents a straightforward use of those technologies together with the experiments' software frameworks.
A future client-server approach to data access could fill the current gap and let developers build new VR and AR tools targeting HEP physicists as end users.

\hypertarget{mobile-tech}{%
\subsubsection{Mobile technologies}\label{mobile-tech}}

Portability and simplicity of usage are the strong points of mobile devices. More than as ``mobile'' devices, smartphones,
tablets and ultrabooks can be considered, as devices ``close to people''. As such, the usage of such devices
should be exploited more in the final steps of the visualization chain, where heavy batch data processing is not needed. For instance,
their usage should be leveraged for the production and visualization of event displays.
Ideally, a user should be able to easily retrieve interesting events from the experiment and interactively visualize them on all
kinds of devices.

That is why we strongly promote the usage of the server-client architecture described and supported in this paper in Section~\ref{client-server}
and the new data access patterns presented and supported in the the HSF \textit{Data Access and Management} Community White Paper~\cite{HSF-CWP-2017-04}. %%% TODO: Check the link to the DOMA paper when out!
This would open up new possibilities for interactive visualization on mobile devices: it would let visualization clients running on mobile devices
connect to server tools running in the experiment's framework to easily and interactively retrieve the desired data.

It is worth noting that in other areas of science, for instance in astronomy, researchers have worked to facilitate data
access and to migrate to more standard data formats. This allowed for the possibility of having data visualization tools
on mobile devices, in addition to desktop and laptop machines. Moreover, this not only helped the researchers in accessing and visualizing
their data, but it also helped in making science accessible by the public, having eased the development of programs used in
Outreach and Education activities and events. It is true that HEP data are usually much more complex than astronomy data, and so it
will be harder to achieve, but we think that an effort in simplifying the access to experimental data would be worthwhile.

Therefore, the leverage of the usage of mobile devices in HEP adds a strong point to the development and the support of common
client-server tools and data exchange formats among HEP experiments in the near future.

\hypertarget{multi-user}{%
\subsubsection{Multi-user applications}\label{multi-user}}

Nowadays multi-user technology is used in many applications: for example in GoogleDocs, where many users can simultaneously
interact with the same document. What we would like to provide is multi-user support for visualization to let several users
explore and interact with events at the same time. Beside being a useful feature for expert users (for example, when asking for advice on the visualization of a piece of detector to another person at a distant institute), it could be important for
outreach and education activities, where students and other people could interact together with an event display, or for virtual guided tours.
Game engines offer multi-user support natively, thus we could starting explore their usage.
An example of such collaborative features in a 3D environment is integrated in the Med3D visualization framework~\cite{Bohak2017}.

\hypertarget{sharing-knowledge}{%
\section{Sharing knowledge and fostering collaboration}\label{sharing-knowledge}}

During the kick-starter meetings and the different workshops organized to start and develop the present Community White Paper,
the whole HSF Visualization Working Group has agreed on the importance of sharing knowledge among the whole HEP community,
as well as best practices and know-how. Too often, in fact, solutions and tools developed for one HEP experiment are
not sufficiently advertised to the rest of the HEP Visualization community, with the result that the community base knowledge
is fragmented and not efficiently exploited.

The focus of this Working Group in fact is not limited to the preparation of this white paper. Instead, a longer term collaboration
among the experiments is foreseen, in order to collaborate on common visualization projects.
To foster collaboration and sharing, we agree on following courses-of-action which are described below.

\hypertarget{workshop}{%
\subsection{Yearly workshop}\label{workshop}}

On March 2017 the first HSF Visualization Workshop was organized at CERN to let all the experts from the different experiments and projects
show their work and share their solutions. In addition, external experts from industry were invited to present the latest
advancements in the field and best practices \cite{HSFViz2017}.

It was the first topical workshop focused on HEP Visualization in many years. Many HEP
experiments and communities showed their latest developments. Given the high number of presentations, a second mini-workshop~\cite{HSFViz2017Mini} has been organized as a follow-up, to let the remaining communities to present their work.

The Working Group agreed on the importance of meeting to share findings, knowledge and solutions; and it was decided to
try to organize a topical workshop on HEP visualization once per year.

An important point was raised while organizing the first Workshop: other scientific fields have visualization and
graphics needs similar to HEP, for example Geophysics. In future workshops we will try to
have presentations from other communities as well, in order to try to foster friendly and fruitful collaborations
which could benefit the whole scientific community.

\hypertarget{repo}{%
\subsection{Code repository}\label{repo}}

The HSF Visualization Working Group also agreed on the importance of fostering collaborative work.
As a start a new dedicated project has been created within the HSF GitHub repository~\cite{HSFVizRepo}.
It is intended to be the space where members of the Working Group can share their work-in-progress studies and their solutions, and where
community-driven projects will be stored.

\hypertarget{roadmap}{%
\section{Roadmap}\label{roadmap}}

\hypertarget{one-year}{%
\subsection{One year}\label{one-year}}

In the first year the Visualization Working Group will work on defining R{\&}D projects, based on the key
points and ideas discussed in this community white paper.

The main goal will be developing techniques and tools which let visualization applications and event displays be
less dependent on specific experiments’ software frameworks, leveraging the usage of common packages and common data formats.

In a first phase, the community will identify common objects and will agree on common definitions, as described in
section~\ref{common-format}. Then, a common data exchange format, either based on custom data formats or, if possible, on
open standards, will be designed by the community. After that, exporters and interface packages would be designed
as bridges from the experiments’ frameworks, which are needed to access data at a high level of detail, to the common packages.

\hypertarget{three-year}{%
\subsection{Three years: ATLAS and CMS Computing TDRs}\label{three-year}}

In the second and third year the Visualization WG will work on designing and building demonstrators to show the feasibility of
the community-driven best practices and tools. The goal will be to get a final design of those tools, to be included in the development
plans of the different experiments. Moreover, the WG will work towards a more convenient access to geometry and event data.
In collaboration with the HSF \textit{Data Access and Management} working group, an API or a service to get streamed event data would be designed.

\hypertarget{five-year}{%
\subsection{Five years: Towards HL-LHC}\label{five-year}}

In the fourth and fifth year, the focus will be on developing the actual community-driven tools, to be used by the experiments
for their visualization needs in production.

The goal will be the usage of the community-developed tools within the experiments’ visualization applications; and
perhaps the usage of a simplified data access, but that depends on the actual feasibility, which will be established
after an initial study.

\hypertarget{conclusions}{%
\section{Conclusions}\label{conclusions}}

Modern and modular visualization tools, which will feature simplified data access and retrieval as well, would leverage
the accessibility, letting end users exploit all the possibilities offered by modern visualization solutions, without the need
of running them on specific platforms, running them within the experiments’ software frameworks, or being bound to specific solutions.
And a better experience will reflect to a better usage, which will positively affect the usage of such visualization tools for
detector development and simulation of new experiments, as well as the data analysis and the upgrade studies of the current ones.

In the end, common community-driven tools will let users of all experiments make use of the latest and best tools, while sharing the development,
the maintenance and the workload among all the experiments.

\hypertarget{acknowledgements}{%
\section{Acknowledgements}\label{acknowledgements}}

The authors would like to thank Guy Barrand \textit{(Geant4, LSST, LAL)} for the fruitful discussions and input about what is done in astronomy in terms of data access and visualization on mobile platforms.

\sloppy
\raggedright
\clearpage
\printbibliography[title={References},heading=bibintoc]

\end{document}